\documentclass[preprint, 3p, table, xcdraw, authoryear]{elsarticle}

\usepackage{lineno}
\modulolinenumbers[1]

\usepackage{setspace}
\journal{Journal of \LaTeX\ Templates}
\makeatletter
\def\ps@pprintTitle{%
    \let\@oddhead\@empty
    \let\@evenhead\@empty
    \def\@oddfoot{\footnotesize\itshape
         {Preprint} \hfill\today}%
    \let\@evenfoot\@oddfoot
    }
\makeatother

\usepackage{doi}
\usepackage{booktabs}
\usepackage{multirow}
\usepackage{amssymb}
\usepackage{amsmath}

\usepackage{hyperref}
\usepackage{orcidlink}

\newcommand{\Lagr}{\mathcal{L}}

\usepackage{hyperref}
\hypersetup{
  colorlinks   = true, 
  urlcolor     = blue, 
  linkcolor    = blue, 
  citecolor   = blue 
}

\newcommand{\beginsupplement}{%
        \setcounter{table}{0}
        \renewcommand{\thetable}{S\arabic{table}}%
        \setcounter{figure}{0}
        \renewcommand{\thefigure}{S\arabic{figure}}%
        \renewcommand{\thesubsection}{\Alph{subsection}}
     }

\begin{document}

\begin{frontmatter}

  \title{A causal intervention framework for synthesizing mobility data and evaluating predictive neural networks}

  \author[ikg]{Ye Hong\corref{correspondingauthor}}
  \cortext[correspondingauthor]{Corresponding author}
  \ead{hongy@ethz.ch}

  \author[ikg]{Yanan Xin}
  \ead{yanxin@ethz.ch}

  \author[sdsc]{Simon Dirmeier}
  \ead{simon.dirmeier@sdsc.ethz.ch}

  \author[sdsc,iml]{Fernando Perez-Cruz}
  \ead{fernando.perezcruz@sdsc.ethz.ch}

  \author[ikg]{Martin Raubal}
  \ead{mraubal@ethz.ch}

  \address[ikg]{Institute of Cartography and Geoinformation, ETH Zurich}
  \address[sdsc]{Swiss Data Science Center, ETH Zurich and EPFL}
  \address[iml]{Institute for Machine Learning, Department of Computer Science, ETH Zurich}

  \begin{abstract}
    Deep neural networks are increasingly utilized in mobility prediction tasks, yet their intricate internal workings pose challenges for interpretability, especially in comprehending how various aspects of mobility behavior affect predictions.
    This study introduces a causal intervention framework to assess the impact of mobility-related factors on neural networks designed for next location prediction -- a task focusing on predicting the immediate next location of an individual.
    To achieve this, we employ individual mobility models to synthesize location visit sequences and control behavior dynamics by intervening in their data generation process.
    We evaluate the interventional location sequences using mobility metrics and input them into well-trained networks to analyze performance variations.
    The results demonstrate the effectiveness in producing location sequences with distinct mobility behaviors, thereby facilitating the simulation of diverse yet realistic spatial and temporal changes.
    These changes result in performance fluctuations in next location prediction networks, revealing impacts of critical mobility behavior factors, including sequential patterns in location transitions, proclivity for exploring new locations, and preferences in location choices at population and individual levels.
    The gained insights hold value for the real-world application of mobility prediction networks, and the framework is expected to promote the use of causal inference to enhance the interpretability and robustness of neural networks in mobility applications.
  \end{abstract}


\end{frontmatter}


\section{Introduction}
Accurate individual mobility prediction plays a pivotal role in popularizing emerging mobility services~\citep{ma_individual_2022} and serves as a crucial backbone for various intelligent transport system functionalities~\citep{tang_delay_2019}.
Despite its application potential, predicting individual mobility remains challenging due to the complexity of mobility patterns influenced by diverse behavioral factors and contexts~\citep{Hong_context_2023}.
The intricacies in modeling these spatiotemporal dependencies generally hinder the prediction performance of individual mobility~\citep{song_limits_2010,barbosa_human_2018,wiedemann_predicting_2023}.
In recent years, the availability of human digital traces and the advancements in data-driven models, particularly deep neural networks capable of capturing spatiotemporal dynamics, have significantly enhanced mobility prediction ability~\citep{wang_deep_2022}.

Despite their solid predictive performance, modern neural networks often face criticism for their low interpretability~\citep{manibardo_deep_2022, pappalardo_future_2023}, referring to the degree to which humans can comprehend the decision-making process of a model.
These networks are commonly regarded as ``black boxes'' because reconstructing the reasoning behind a particular prediction is challenging.
In mobility prediction, the lack of interpretability leads to an unclear understanding of the spatiotemporal patterns captured by the network and, more fundamentally, the influence of behavioral factors (such as location preferences and activity radius~\citep{yuan_correlating_2012}) in prediction.
This deficiency negatively affects decision-making, policy design, and the perceived reliability and trustworthiness among practitioners~\citep{huang2020survey}, thereby impeding the seamless integration of mobility prediction networks into real-world applications~\citep{koushik_machine_2020}.
Furthermore, the scarcity of publicly available individual mobility datasets, primarily due to the privacy-sensitive nature of personal mobility~\citep{wiedemann_influence_2023}, leads to a lack of comparability between existing and newly developed prediction models~\citep{graser_deep_2023}.
Prediction networks are evaluated using datasets that include varying numbers and types of participants, along with differing tracking durations, representing diverse snapshots of the possible mobility behavior~\citep{kulkarni201920}.
Hence, a comprehensive analysis connecting behavior dynamics with prediction performance is imperative to establish benchmark data specifications for evaluating modern neural networks employed in mobility studies.

In addition, establishing the behavior and performance connection assists in evaluating the robustness of these networks when confronted with unforeseen inputs.
The optimization of neural networks requires a training dataset, making their performances heavily dependent on the quality and representativeness of this data~\citep{yin_deep_2022}.
However, mobility behavior evolves dynamically over space and time due to internal needs (e.g., behavioral exploration~\citep{hong_conserved_2023}) and environmental factors (e.g., land use~\citep{acheampong_towards_2018}).
Consequently, the mobility data encountered during application often reflects different behavior than the training data, leading to a discrepancy known as domain shift~\citep{he_what_2020}.
Enhancing our understanding of performance under various shift scenarios is essential to assessing reliability when applying these networks across diverse geographic regions or time periods. Yet, this relationship remains predominantly unexplored.

Causal intervention offers a promising tool for generating data from diverse environments, enabling the assessment of neural network robustness and providing human-friendly causal explanations for these interventions~\citep{xin_vision_2022}.
Building upon its advantages, we present a framework for systematically evaluating the impact of mobility behaviors on prediction networks.
Specifically, this framework utilizes individual mobility models to generate mobility traces, and employs causal intervention strategies in the data generation process, allowing for flexible modifications of the defined mobility behavior.
We subsequently assess the performance of trained neural networks on these synthetic traces for mobility prediction.
Our study focuses on next location prediction, which aims to forecast an individual's immediate next location based on their mobility history.
The interventions simulated realistic spatial and temporal changes in mobility patterns, leading to performance fluctuations in prediction networks that reflect their robustness when confronted with domain shifts.
This framework facilitates assessing the impact of behavioral factors and benchmarking mobility prediction networks, with practical applications for evaluating network performances and transferring these networks across environments.
In short, our contributions are summarized as follows:
\begin{itemize}
  \item We introduce a framework to assess the robustness of mobility prediction networks through causal intervention. This framework enables direct control over behavioral dynamics, quantifying ensuing mobility patterns, and evaluating their influence on network performance.
  \item We use this framework to simulate spatiotemporal shift scenarios, demonstrating its effectiveness in benchmarking mobility datasets, identifying performance degradation during real-world applications, and improving the interpretability of prediction networks.
  \item We open-source the framework, enabling straightforward utilization and flexible customization of its components\footnote{The source code is available at \url{https://github.com/irmlma}}.
\end{itemize}

\section{Related work}\label{sec:related}

\subsection{Mobility behavior and its impact on mobility prediction}

Research on individual mobility behavior, with an emphasis on spatiotemporal patterns of activities and trips, has consistently been at the forefront of mobility studies~\citep{chen_promises_2016}, propelling theoretical and methodological advancements within the activity-based analysis framework~\citep{schonfelder_urban_2016}.
In this framework, trips are perceived as an induced demand resulting from the necessity to engage in activities at distinct spatial locations~\citep{axhausen_activitybased_1992}, making location selection a pivotal aspect in comprehending individual mobility behavior~\citep{sener_accommodating_2011}.
Studies have consistently shown that location choices not only vary across populations (inter-person variability)~\citep{martin_graph_2023, ji_rethinking_2023} but also undergo constant changes over time (intra-person variability)~\citep{susilo_repetitions_2014, hintermann_impact_2023}.
Furthermore, these decisions are frequently influenced by external factors, some of which, when encountered, can lead to sudden structural changes, such as residential relocation~\citep{ramezani_residential_2021} and large-scale crises~\citep{kellermann_mobility_2022}.
These empirical insights emphasize the need to study activity location choices from a dynamic perspective~\citep{shou_similarity_2018}, rather than approaching them solely with static mobility snapshots.

Considering the central role of activity, predicting activity location is a crucial component in mobility ahead planning and optimization~\citep{ma_individual_2022}, and has garnered widespread attention~\citep{luca_survey_2021}.
Accurately inferring the next location is influenced not only by the capability of the employed model (e.g., neural networks) but also by individuals' behavioral patterns.
In a pioneering effort to link mobility behavior with location prediction, \citet{song_limits_2010} proposed entropy as a measure for the theoretical mobility predictability based on visited location sequences of individuals.
Their study revealed remarkably consistent predictability across individuals, peaking at approximately 93\% in the tested dataset, reflecting inherent travel patterns across various demographic attributes~\citep{song_limits_2010}.
Although studies have highlighted their strong correlation with the actual prediction performance~\citep{lu_approaching_2013}, entropy-related predictability measures remain challenging to interpret~\citep{teixeira_deciphering_2019}.
Recent research has started to represent mobility behavior with more straightforward metrics, such as routine and novelty components~\citep{teixeira_estimating_2021}, usages of transport services~\citep{xu_understanding_2022} and recurring patterns in daily mobility~\citep{Hong_context_2023}, and analyze its impact on prediction performance.
Nevertheless, our comprehension of these connections is still in its early stages, relying on observed mobility behavior from tracking datasets, which often exhibit limited behavior variability and biased behavior distribution.

\subsection{Causal intervention and its application in robustness assessment}

The field of causal modeling, which investigates causal relationships, provides a systematic framework for generating mobility traces that depict specific behaviors.
Establishing a structural causal model (SCM) is essential to describe the causal mechanisms within a system comprising various interconnected factors.
In this context, the data generation process is viewed as a causal process, where the input variables of the generation model are considered the causes of the output, with their relationships modeled through SCM~\citep{rahimi_counterfactual_2023}.
With access to a particular SCM, we can perform interventions on individual factors or combinations thereof, effectively disentangling intricate interactions among variables by unilaterally adjusting the value of a single variable and observing its impact on the generated data~\citep{pearl_book_2018}. 
Utilizing causal interventions to generate controlled results has diverse applications.
For instance, in prognostic research, causal intervention has been applied to forecast risks associated with various medical treatments~\citep{van2019eliminating}, examine the impact of therapy on the cognitive development of prematurely born infants~\citep{silva2016observational}, and investigate the effects of demographic variations on brain structure using counterfactual samples~\citep{pawlowski2020deep}. 
Similar strategies have also been applied in climate and earth science to unravel the complexities of dynamic spatiotemporal processes~\citep{li_big_2023, runge_causal_2023}.

Causal intervention methods have demonstrated potential in evaluating and enhancing the robustness of neural networks~\citep{scholkopf_toward_2021, moshkov_learning_2024}, where robustness refers to the degree of performance variations under domain shift~\citep{zhou_domain_2023}.
In this study, we used SCM to abstract the data generation process of mechanistic mobility simulators, which incorporate parameters representing individual behavior to synthesize mobility traces.
Noteworthy simulators include the exploration and preferential return (EPR) model~\citep{song_modelling_2010}, the location attractiveness model~\citep{yan_universal_2017}, and the container model~\citep{alessandretti_scales_2020}.
These models have significantly advanced in replicating high-level spatiotemporal patterns of individual movements~\citep{pappalardo_future_2023} and are increasingly applied in large-scale simulations \citep{xu_planning_2018, barbosa_human_2018}.
Representing these models as SCMs enables the generation of both in-distribution and out-of-distribution (OoD) data by intervening in input variables and adjusting their strengths.
For example, increasing the tendency of individuals to explore new locations in the EPR model allows us to generate mobility data from a more explorative population.
As a result, various interventions in mobility models produce a spectrum of data, each reflecting a realistic domain shift, enabling us to assess the robustness of prediction networks in different scenarios.

\section{Methodology}\label{sec:method}

\begin{figure*}[!ht]
  \centering
  \includegraphics[width=1.0\textwidth]{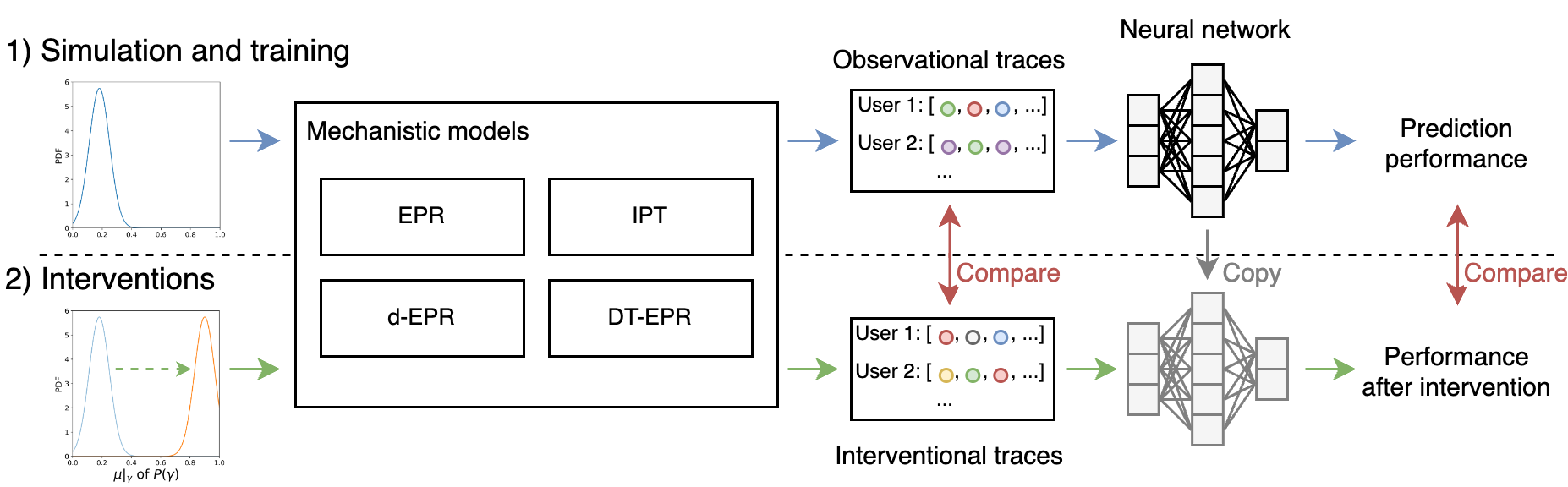}
  \caption{Framework for evaluating the robustness of prediction networks through causal interventions. We generate location sequences from mechanistic models and feed them into prediction networks to evaluate the prediction performance (blue arrows). This process is repeated for interventional location sequences, obtained by modifying the distribution of behavioral parameters (green arrows). The differences in mobility patterns and prediction performances are compared to assess intervention strengths and network robustness (red arrows).}
  \label{fig:overview}
\end{figure*}

The overall pipeline for assessing the robustness of prediction networks is illustrated in Figure~\ref{fig:overview}.
We start by introducing mechanistic generative models for synthesizing individual location sequences (\(\S\)\ref{sec:generative_models}). These models incorporate parameters to replicate real-world observational mobility behavior.
Subsequently, we perform interventions by modifying the parameters, thereby manipulating mobility behaviors and generating new interventional location sequences (\(\S\)\ref{sec:intervention_design}).
Lastly, we train mobility prediction networks using observational location sequences and evaluate them on interventional sequences (\(\S\)\ref{sec:predictive_models}). Prediction performance changes reflect these networks' robustness when behavioral interventions are introduced.
In this study, the term \textit{location} refers specifically to a geographical place where individuals engage in activities. This excludes other points, such as waypoints on roads (e.g., GPS recordings) or intermediate stops without meaningful activity (e.g., waiting for a bus).
We provide a more detailed description of each module in the following sections.

\subsection{Individual mobility models}\label{sec:generative_models}

\begin{figure*}[!ht]
  \centering
  \includegraphics[width=0.6\textwidth]{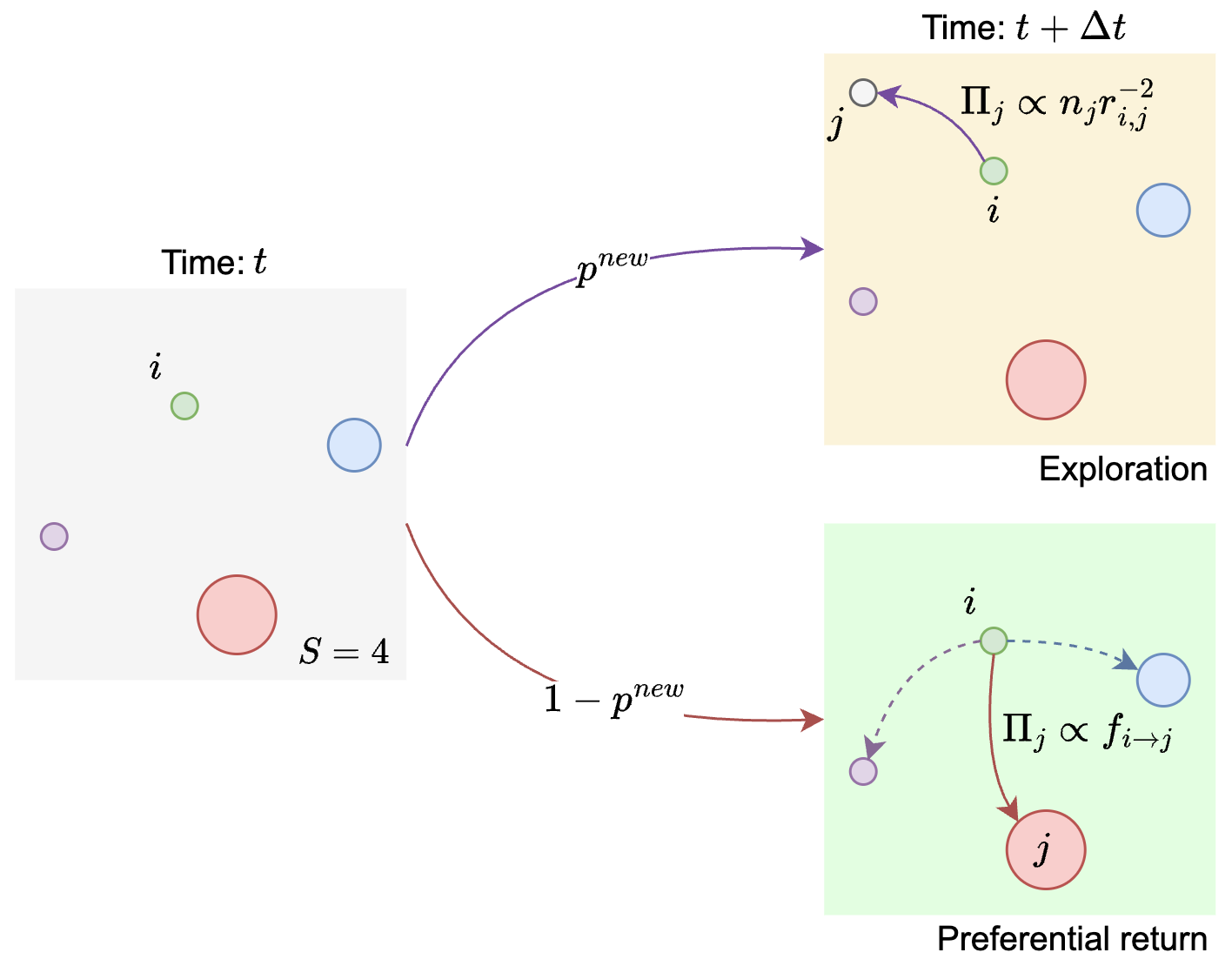}
  \caption{Mechanistic generative model DT-EPR. The individual at location $i$ visited $S=4$ locations with a frequency proportional to the size of the location circle at time $t$. At time $t+\Delta t$, the individual chooses to either explore a new location with probability $p^{new}$, where the next location $j$ will be chosen based on its population attractiveness $n_j$ and the travel distance $r_{i,j}$ (d-EPR mechanism; upper panel), or return to a previously visited location with complementary probability $1-p^{new}$, where the location probability $\Pi_j$ is proportional to the empirical visit frequency from $i$ (IPT mechanism; lower panel). Figure adapted from~\citet{song_modelling_2010}.}
  \label{fig:epr_detail}
\end{figure*}

Individual mobility models generate realistic movement trajectories based on a predefined set of behavioral parameters, allowing for direct control over the mobility behavior of the generated population.
We start with the EPR model~\citep{song_modelling_2010} as our baseline SCM and additionally introduce two EPR-based generative models, namely density (d)-EPR~\citep{pappalardo_returners_2015} and individual preferential transition (IPT)~\citep{zhao_characteristics_2021}. We finally propose the density transition (DT)-EPR model that combines the advantages of d-EPR and IPT to obtain more realistic mobility traces. Figure~\ref{fig:epr_detail} delineates the mechanisms of DT-EPR and its connection with d-EPR and IPT models.

EPR generates sequences of location visits for individuals and reproduces scaling laws for distinct location numbers and their visitation frequency over time~\citep{ barbosa_human_2018}.
The core of EPR is the introduction of two competing mechanisms, exploration and preferential return, into classical random-walk models~\citep{brockmann_scaling_2006}. These two mechanisms account for the tendency of individuals to return to previously visited locations~\citep{song_modelling_2010}.
Specifically, observing an individual at location $i$ at time $t$, the model assumes that the individual will change their location after a waiting time $\Delta t$, where $\Delta t$ is sampled from its distribution $P(\Delta t)$.
The individual chooses to explore a previously unvisited location with probability $p^{new}_{t+\Delta t}$:
\begin{linenomath}
  \begin{equation}\label{eq:exploration}
    p^{new}_{t+\Delta t}=\rho S_{t}^{-\gamma}
  \end{equation}
\end{linenomath}
where $0 < \rho \leq 1$ and $\gamma \geq 0$ are parameters that control the exploration tendency and $S_{t}$ denotes the number of distinct location visited until time $t$. During this process, a new location is determined by sampling a moving distance $\Delta r$ from the jump length distribution $P(\Delta r)$, with the moving direction chosen uniformly at random. After the move, the number of visited locations increases from $S_{t}$ to $S_{t}+1$.
Besides exploring a new location, the individual could return to a visited location with complementary probability $1-p^{new}_{t+\Delta t}$. In this case, the probability of moving to a location $j$, denoted as $\Pi_j$, is proportional to the number of previous visits to $j$, i.e., $\Pi_j \propto f_j$, where $f_j$ is the visitation frequency of $j$.

Later variants of EPR have modified the search for locations to replicate spatial patterns.
One such variant is the d-EPR model, which addresses the insufficient reproduction of the evolution of the radius of gyration~\citep{pappalardo_returners_2015}.
A population attractiveness factor is assigned to each location to model the tendency to visit popular locations (Figure~\ref{fig:epr_detail} upper panel).
In the model, the probability $\Pi_j$ of selecting location $j$ during exploration depends on its travel distance and attractiveness:
\begin{linenomath}
  \begin{equation}\label{eq:explore_location}
    \Pi_j \propto n_jr_{i,j}^{-2}
  \end{equation}
\end{linenomath}
where $r_{i,j}$ is the distance between the current location $i$ and the new location $j$, and $n_j$ denotes the attractiveness, quantified as the empirical visits by all individuals to location $j$.
Furthermore, a subsequent study has identified that individuals can travel arbitrarily large distances during preferential return, as the location selection of EPR is based solely on empirical visit frequency~\citep{zhao_characteristics_2021}.
To address this limitation, \citet{zhao_characteristics_2021} proposed the IPT model that constrains the preferential return to be conditioned on the current location, i.e., imposing a first-order Markov process on location choices (Figure~\ref{fig:epr_detail} lower panel).
The visit probability of location $j$ is defined as proportional to the number of previous visits from the current location $i$ to $j$:
\begin{linenomath}
  \begin{equation}\label{eq:pr}
    \Pi_j \propto f_{i\rightarrow j}
  \end{equation}
\end{linenomath}
In practice, a personalized Markov transition matrix containing $f_{i\rightarrow j}$ for each location pair is initialized from empirical location visits and updated during the generation process.

We combine the exploration mechanism of d-EPR and the preferential return mechanism of IPT to introduce a new EPR-based model, which we refer to as DT-EPR (Figure~\ref{fig:epr_detail}). Consequently, DT-EPR inherits the capacity to capture both population attractiveness and individual preferences in location choices.
As a result, for each individual $u^i$ in the user set $\mathcal{U} = \left \{ u^{1}, ..., u^{\left |\mathcal{U}  \right |} \right \}$, DT-EPR generates a time-ordered trajectory $T^i = \left (L_k  \right )_{k=1}^{m_{u^i}}$ composed of $m_{u^i}$ locations visited by $u^i$. A location $L$ contains spatiotemporal information and is represented as a tuple of $L = \langle l, p, t\rangle $, where $l$ is the location identifier, $p = \langle x, y\rangle $ represents spatial coordinates in a reference system, e.g., latitude and longitude, and $t$ is the time of visit. Thus, we construct the location set $\mathcal{O}^i$ containing the known location identifiers for individual $u^i$, and the set $\mathcal{O} = \left \{ \mathcal{O}^1, ..., \mathcal{O}^{\left |\mathcal{U}  \right |} \right \}$ including all locations in the dataset.

\subsection{Intervention design}\label{sec:intervention_design}

We use empirically estimated behavioral parameters for the DT-EPR model, generating \textit{observational} mobility traces.
Since DT-EPR is a parametric simulator, i.e., parameters have real-world interpretations with explainable relationships with other model variables, we can introduce causal interventions to the data-generating process to simulate \textit{interventional} mobility trajectories.
Causal interventions can be interpreted as shifts in the observed mobility patterns, representing various scenarios, such as spatial shifts when certain locations become more or less attractive or temporal shifts in mobility behavior between seasons or years.

Considering the modeling mechanisms of the DT-EPR model, we perform interventions on the following parameters to simulate comprehensive behavior shift scenarios:
\begin{itemize}
  \item The exploration tendency $p^{new}$, affecting whether or not to explore in the next time step (Eq.~\ref{eq:exploration}).
        While the number of distinct locations collectively exhibits sublinear growth~\citep{song_modelling_2010}, notable inter-person variability in location choices implies variations in the exploration speed among individuals~\citep{hong_conserved_2023}.
        In EPR-like models, $p^{new}$ is determined by parameters $\rho$ and $\gamma$, independently sampled for each individual from empirical distributions.
        We introduce interventions on $\rho$ and $\gamma$ by altering their distributions, producing pseudo-populations with different exploration behaviors.
        Formally, let $\mathbb{P}_{\mathcal{S}}^{ (\rho, \gamma, p^{new})}$ be the joint probability distribution over random variables $\rho, \gamma, p^{new}$ induced by the SCM $\mathcal{S}$. We intervene on each of these variables, e.g., for the case of $\rho$, we replace the generating mechanism of $\rho$ with a new one $\tilde{f}_\rho$, i.e., $\mathbb{P}_{\mathcal{S}}^{ (\rho, \gamma, p^{new} ) \mid do(\rho = \tilde{f}_\rho(\cdot)) }$. The intervention $do(\rho = \tilde{f}_\rho(\cdot))$ induces a new probability distribution $\mathbb{P}_{\tilde{\mathcal{S}}}^{ (\rho, \gamma, p^{new} ) }$ with modified structural equations that we use for sampling interventional data.
        Additionally, we perform hard interventions on $p^{new}$ by fixing its value to a constant. This intervention removes the time-dependent modeling of location exploration, falling back to the assumption in the L\'evy flight model~\citep{brockmann_scaling_2006}. Employing the same formalism as above, a hard intervention puts point mass on a specific outcome of a structural equation, i.e., $do(p^{new} = a)$, where $a$ is a constant value.
  \item The population attractiveness $n$, affecting location choices during exploration (Eq.~\ref{eq:explore_location}).
        We manipulate location attractiveness to simulate changes in the population's spatial preferences. To retain location visitation characteristics, we randomly shuffle empirical visit numbers for a group of locations.
        The strength of the intervention can be controlled by adjusting the group of locations, e.g., including more locations in the shuffling process introduces a more substantial intervention.

  \item The empirical individual preference $f$, affecting location choices during preferential return (Eq.~\ref{eq:pr}).
        As individual mobility behavior is dynamic and varies considerably over time, we introduce interventions by manipulating the Markov transition matrix for each individual.
        This is achieved by shuffling the empirical visit numbers for a group of locations, which maintains the overall number of visits while altering the choice probabilities for each location.
        The strength of the intervention is controlled by selecting the location group to include in the shuffling process.
\end{itemize}

For each intervention, the DT-EPR model generates interventional mobility trajectories $\tilde{T^i} = \left (L_k  \right )_{k=1}^{m_{u^i}}$ for individual $u^i \in \mathcal{U}$, which share an identical data format as the observational mobility traces $T^i$.

\subsection{Next location prediction networks}\label{sec:predictive_models}

To assess the influence of causal interventions, i.e., the impact of changes in mobility behavior, we evaluate the predictive capability of a neural network trained on observational data but tested on interventional data.
We choose next location prediction as the application task.
Practically, consider a sub-sequence $\left (L_k  \right )_{k=m}^{n} \in T^i$ visited by individual $u^i$ in a time window from time step $m$ to $n$, the goal is to predict the location the same individual will visit in the next time step, i.e., the location identifier $l_{n+1} \in \mathcal{O}$.
While conventional approaches relied on Markov models and matrix factorization methods, recent years have witnessed a growing adoption of neural networks~\citep{luca_survey_2021}.
In the following, we present a typical pipeline for applying these networks to next location prediction, including generating feature embedding, designing the prediction network, and defining a loss function for parameter optimization.

An effective location prediction method starts by selecting and modeling the information in historical sequences.
We include location identifiers, the time for location visits, and information regarding the individual who conducted the visit.
To represent these features, we introduce embedding layers that utilize parameter matrices to map the original variables to real-valued embedding vectors.
For each location $L_k$, vector representations of its location identifier $l_k$ and time of arrival $t_k$ are obtained as follows:
\begin{linenomath}
  \begin{equation}
    e^{l}_{k} = h^l(l_{k}; \boldsymbol{W}^l) \qquad e^{t}_{k} = h^t(t_{k}; \boldsymbol{W}^{t})
  \end{equation}
\end{linenomath}
where $e^{l}_{k}$ and $e^{t}_{k}$ are the respective embedding vectors, $h(\cdot; \cdot)$ denotes the embedding operation, and $\boldsymbol{W}$ terms are the learned parameter matrices during training.
To capture different periodicity, we separately embed the minute, the hour, and the day of the week from the visit time $t_k$.
The overall embedding vector $e^{all}_{k}$ is obtained by adding the location and temporal embedding vectors: $e^{all}_{k} = e^{l}_{k} + e^{t}_{k}$.
Additionally, we represent the individual \(u^i\) that conducted the travel into a vector \(e^{u^i}\) through a user embedding layer, i.e., $e^{u^i} = h^u({u^i}; \boldsymbol{W}^{u})$.
Therefore, we obtain the overall embedding vector $e^{all}_{k}$ that encodes spatiotemporal features at each time step and the user embedding vector $e^{u^i}$ corresponding to each location sequence.

A location prediction network aims to learn the transition patterns between historical location visits in order to predict the next possible location.
Without loss of generality, this process can be viewed as obtaining a (highly non-linear) mapping $g$ with learnable parameters $\boldsymbol{W}^{g}$ between the visit sequence $\left (e^{all}_k \right )_{k=m}^{n}$, user information $e^{u^i}$ and the ground-truth next location $l_{n+1}$, i.e., $l_{n+1} = g(\left (e^{all}_k \right )_{k=m}^{n}, e^{u^i}; \boldsymbol{W}^{g})$.
When a new visit sequence is observed, the mapping $g$ is used to infer the predicted next location $\hat{l}_{n+1}$.
Various sequential modeling methods can be employed to approximate $g$, among which Long Short-Term Memory (LSTM)~\citep{hochreiter_long_1997} and Multi-Head Self-Attention (MHSA)-based~\citep{vaswani_attention_2017} networks have gained considerable attention in the field.
We implement LSTM and MHSA-based networks for next location prediction and present details of their architectures and realizations in Appendix~\ref{app:network}.

In practice, for efficient parameter optimization, the prediction network outputs visiting probabilities instead of directly predicting the next location:
\begin{linenomath}
  \begin{equation}
    \label{equation:fc}
    P(\hat{l}_{n+1}) = \text{Softmax} (g(\left (e^{all}_k \right )_{k=m}^{n}, e^{u^i}; \boldsymbol{W}^{g}) )
  \end{equation}
\end{linenomath}
where $P(\hat{l}_{n+1}) \in [0, 1]^{\left | \mathcal{O}  \right |}$ contains visit probabilities of all locations at the next time step, and Softmax denotes the softmax operation, ensuring $P(\hat{l}_{n+1})$ is a valid probability distribution.
With access to the ground truth next location $l_{n+1}$ during training, we can regard it as a multi-class classification problem, such that the network can be optimized using the multi-class cross-entropy loss $\Lagr$:
\begin{linenomath}
  \begin{equation}
    \label{equation:loss}
    \Lagr = -\sum_{k=1}^{\left |\mathcal{O}  \right |}P(l_{n+1})^{(k)}\log(P(\hat{l}_{n+1})^{(k)})
  \end{equation}
\end{linenomath}
where $P(\hat{l}_{n+1})^{(k)}$ represents the predicted probability of visiting the $k$-th location (the $k$-th entry in $P(\hat{l}_{n+1})$) and $P(l_{n+1})^{(k)}$ is the one-hot represented ground truth, i.e., $P(l_{n+1})^{(k)}=1$ if and only if the $k$-th location corresponds to the ground truth next location.

\section{Dataset and Experiment}\label{sec:data}

\subsection{Movement data and preprocessing}

Individual mobility models rely on behavioral parameters calibrated from real-world data to depict mobility behavior.
Smartphone-based travel surveys have emerged as a promising method for acquiring high-quality spatiotemporal travel behavior data~\citep{harding_are_2021} and are increasingly used for monitoring and studying travel patterns \citep{joseph_measuring_2020, rout_using_2021, mandal_exploring_2023}.
To estimate these behavioral parameters, we leverage a smartphone-based travel survey conducted by the Swiss Federal Railways (SBB), known as the SBB Green Class (GC) E-Car pilot study, which aimed to assess the impact of a Mobility-as-a-Service (MaaS) offer on travel behavior~\citep{martin_begleitstudie_2019}.
The pilot study yielded a large-scale longitudinal GNSS tracking dataset from 139 participants located in Switzerland, spanning from November 2016 to December 2017.
As part of the data collection process, participants were asked to install a commercial application on their smartphones, which continuously recorded their whereabouts from GNSS signals.
By analyzing motion measurements such as speed and acceleration, the application pre-processed the raw traces to identify two types of mobility events: \textit{stay points} representing areas where users were stationary, and \textit{stages} representing continuous movements using a single travel mode.
Post-tracking analysis revealed that the median time between two consecutive GNSS recordings was 13.9 seconds, indicating a high temporal tracking quality of individual movements throughout the study period.

We perform a series of preprocessing steps on the GC dataset.
First, to ensure high-quality tracking data for analysis, we include participants with extended tracking periods (observed for more than 300 days) and high temporal tracking coverage (whereabouts known for more than 60\% of the time).
Then, locations, the basic movement units in individual mobility models, are derived from stay point sequences. We identify \textit{activities} from stay points with durations exceeding 25 minutes. Locations are formed by aggregating activity stay points spatially: we use the DBSCAN clustering algorithm with parameters \(\epsilon = 20\) and \(num\_samples = 1\) from the \textit{Trackintel} library to generate \textit{dataset}-level locations~\citep{Martin_trackintel_2023}.
The longitudinal tracking of the GC study provides long-term observations of participants' daily activity location choices, allowing for precise estimation of behavioral parameters and a sufficiently large location choice set during mobility simulations. Since individuals' tendency to explore new locations diminishes significantly over time~\citep{song_modelling_2010, alessandretti_evidence_2018}, longitudinal observations reduce the occurrences of location exploration, thereby mitigating its impact on location prediction performance~\citep{cuttone_understanding_2018}.

\subsection{Realization of mobility models}\label{sec:realize_model}

The preprocessed GC dataset is used to estimate the behavioral parameters, including the jump length distribution $P(\Delta r)$ for vanilla EPR, as well as the waiting time distribution $P(\Delta t)$ and distributions for exploration tendency ($P(\rho)$ and $P(\gamma)$) for EPR-like models.
Specifically, we consider the log-normal distribution and the power law (including truncated power law) distribution as candidate distributions for determining $P(\Delta r)$ and $P(\Delta t)$~\citep{alessandretti_multi_2017}. We use the functions provided by the \textit{powerlaw} library~\citep{alstott_powerlaw_2014} and evaluate the goodness-of-fit using the Akaike information criterion (AIC) and Akaike weights~\citep{zhao_explaining_2015}. Under AIC, both distributions are best fitted using a log-normal distribution of the probability density function $P(x)= \frac {1}{x\sigma {\sqrt {2\pi }}}\ \exp \left(-{\frac {\left(\ln \left(x\right)-\mu \right)^{2}}{2\sigma ^{2}}}\right)$ with parameters $\mu\vert_{\Delta r} = 7.72$, $\sigma\vert_{\Delta r} = 2.38$, and $\mu\vert_{\Delta t} = 0.75$, $\sigma\vert_{\Delta t} = 1.49$, respectively.
To estimate exploration tendencies $P(\rho)$ and $P(\gamma)$, we calculate the location exploration speed and fit normal distributions across individuals. We obtain $\mu\vert_{\rho} = 0.18 $, $\sigma\vert_{\rho} = 0.07$ and $\mu\vert_{\gamma} = 0.64 $, $\sigma\vert_{\gamma} = 0.16$, respectively. These values are consistent with the numbers presented in~\cite{song_modelling_2010}.

We simulate traces for 800 individuals in the observational dataset and each interventional dataset. For each synthetic individual, we independently sample $\rho$ and $\gamma$ from the distributions $P(\rho)$ and $P(\gamma)$, respectively, and randomly assign them to start at one of their top-5 visited locations empirically observed from GC. The generation process continues until individuals have visited 2000 locations, which approximately matches the tracking length of the GC study.

After obtaining observational location sequences $T^i$ and interventional location sequences $\tilde{T^i}$, we compare their behaviors using mobility metrics:
\begin{itemize}
  \item The real entropy~\citep{song_limits_2010} quantifies the regularity of location visit sequences by considering the order and frequency of location visits. It provides insights into the theoretical predictability of mobility~\citep{Hong_context_2023}.
  \item The mobility motifs~\citep{schneider_unravelling_2013} refer to recurring sub-structures that arise when representing daily location visits as graphs. These motifs represent common mobility patterns shared among individuals, and the proportion of motifs for an individual reflects the prevalence of these shared patterns~\citep{cao_characterizing_2019}.
\end{itemize}

We present additional metrics, including radius of gyration~\citep{gonzalez_understanding_2008}, number of transited location pairs~\citep{zhao_characteristics_2021}, and location visitation frequency~\citep{gonzalez_understanding_2008}, in Appendix~\ref{app:metrics}.

\subsection{Implementation of prediction networks}

Next location prediction networks consider the past seven days as historical length; that is, the goal is to predict the following location an individual will be visiting based on their mobility history in the previous seven days.
We adopt a common problem formulation, where prior mobility knowledge is assumed to be available for each individual. Hence, we partition each dataset (i.e., observational and interventional) into non-overlapping train, validation, and test sets based on time. The splitting ratio is set to 3:1:1, where locations occurring in the first 60\% of days for each individual are assigned to the train set, and the last 20\% are assigned to the test set.
We utilize the observational training set to optimize network parameters, and the corresponding validation set to monitor the network loss and performance during training.
We conduct a grid search to find the optimal hyper-parameters that minimize the network loss on the same validation set.
The considered hyper-parameters, the search ranges, and the final selected values can be found in Appendix~\ref{app:network}.
Finally, we evaluate network performances using the held-out test sets from observational and interventional traces.

We use the following performance metrics to evaluate the network prediction:
\begin{itemize}
  \item Accuracy$@$k (Acc$@$k) measures the correctness by checking if the ground truth location is among the top-$k$ most likely visited locations predicted by the network (in $P(\hat{l}_{n+1})$ from Eq.~\ref{equation:fc}). We report Acc$@$1, Acc$@$5, and Acc$@$10.

  \item F1 score (F1) calculates the harmonic mean of precision and recall, considering the uneven visitation preferences to locations in daily routines. We utilize an F1 score weighted by the visit frequency to emphasize the performance in predicting important locations.

  \item Mean reciprocal rank (MRR) assesses the relevance of the recommended locations and is defined as the average reciprocal rank at which the ground truth location is identified by the network (in $P(\hat{l}_{n+1})$ from Eq.~\ref{equation:fc}).
\end{itemize}

\section{Results}\label{sec:result}

\subsection{Mobility simulation and intervention}

\begin{figure*}[!htbp]
  \centering
  \includegraphics[width=0.95\textwidth]{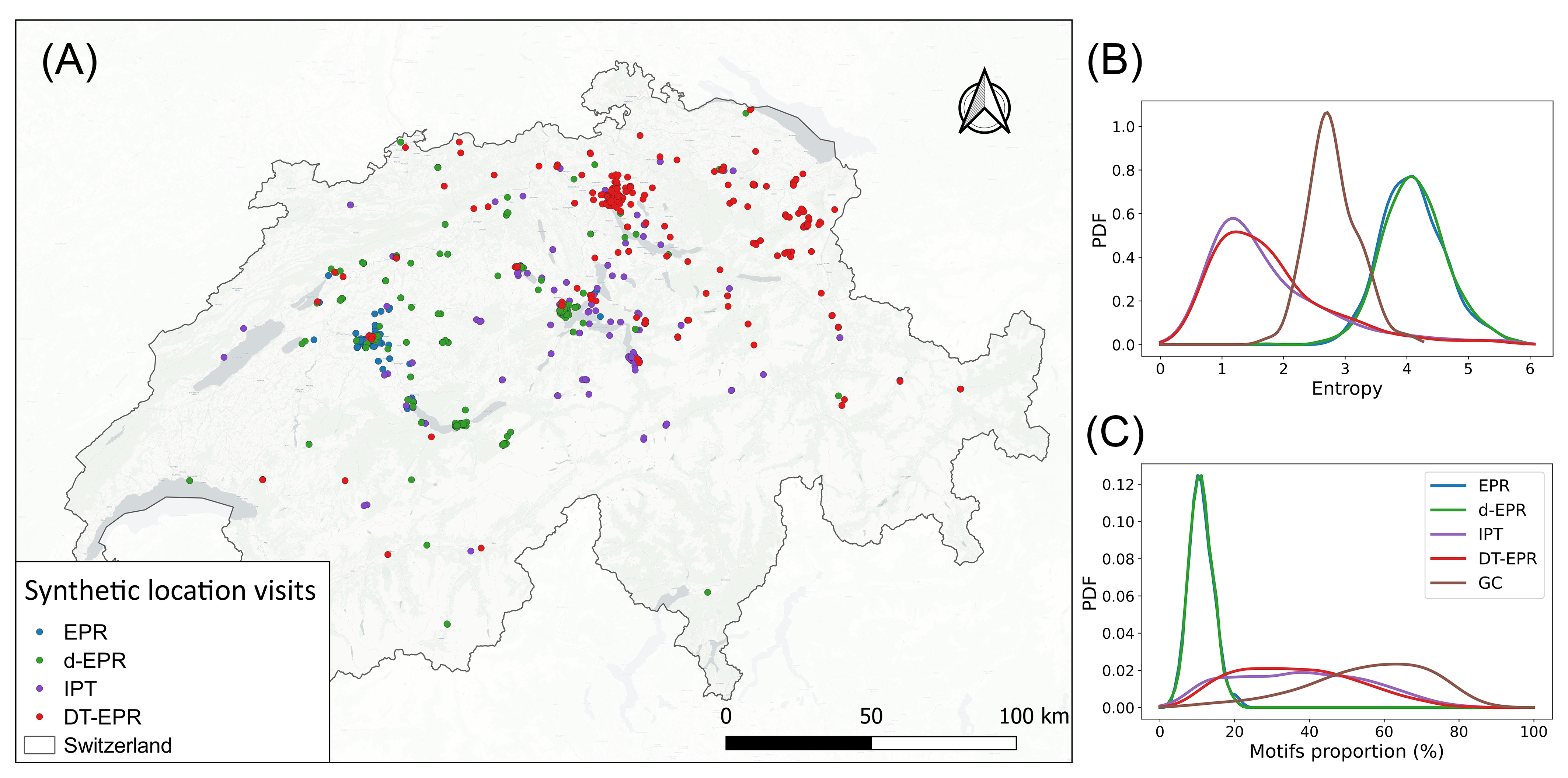}
  \caption{Synthetic location visits generated by EPR-like individual mobility models. (A) Spatial distribution of location visits for an exemplary user. (B) Mobility entropy and (C) motifs proportion distributions of the generated population. Map data ©OpenStreetMap contributors, ©CARTO.}
  \label{fig:demo_generation}
\end{figure*}

\begin{figure}[!htb]
  \centering
  \includegraphics[width=0.8\textwidth]{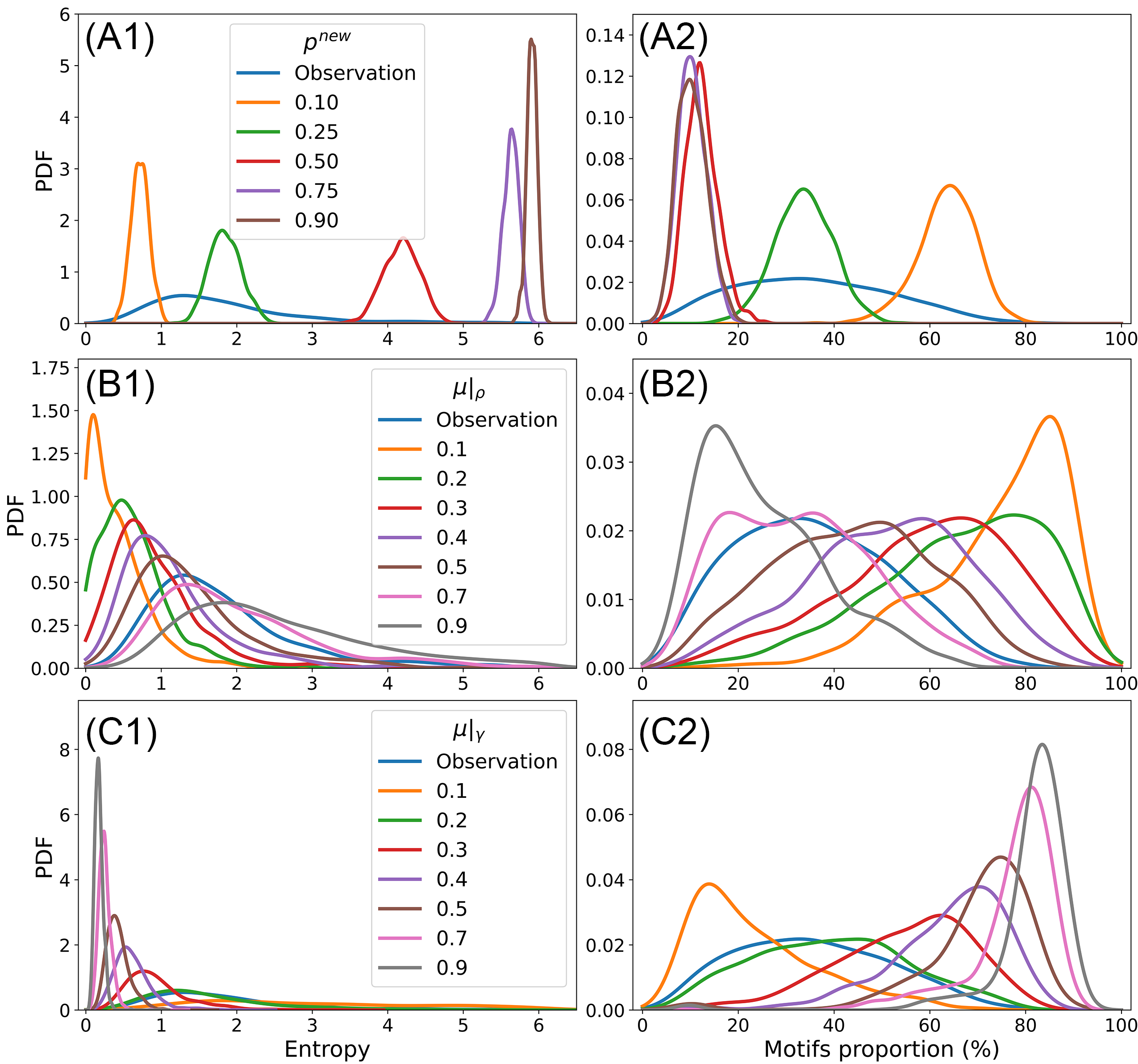}
  \caption{The entropy and motifs proportion distributions of observational and interventional location sequences. We show the metric distributions for (A) hard interventions on $p^{new}$, (B) interventions on $\rho$ by shifting $\mu\vert_{\rho}$ of $P(\rho)$, and (C) interventions on $\gamma$ by shifting $\mu\vert_{\gamma}$ of $P(\gamma)$.}
  \label{fig:generation_metric}
\end{figure}

Leveraging individual models and empirically estimating their behavior parameters, we obtain location traces for synthetic individuals.
The generation result is illustrated in Figure~\ref{fig:demo_generation}, where we map a selected individual's locations and plot the entropy and motifs proportion measures for the entire dataset.
The spatial distribution of location visits (Figure~\ref{fig:demo_generation}A) clearly shows the patterns exhibited by EPR-like models.
The vanilla EPR model assigns visit probabilities based solely on the geographical distance, thus generating location visits clustered around a central area.
While the traces obtained from a d-EPR model are more concentrated in large cities, the locations generated by an IPT model are more dispersed in space, reflecting the empirical knowledge of that specific individual.
Finally, the DT-EPR model combines the advantages of both d-EPR and IPT models, producing traces that prioritize locations within urban areas (population attractiveness) and occasionally occur in distant sites (personal preferences).
Furthermore, we quantitatively assess the regularities of the synthetic sequences at the dataset level (Figure~\ref{fig:demo_generation}B and C).
EPR and d-EPR exhibit the highest entropy and lowest motifs proportion among all mobility models, indicating that their location visit patterns are irregular and hard to predict.
Although the d-EPR model introduces population attractiveness in the exploration phase, it does not alter the process of visiting known locations, which is the primary mechanism underlying mobility's regularity.
On the contrary, DT-EPR incorporates the first-order Markov dependence on location visits, producing the most regular location visitation patterns, as evidenced by the lowest entropy and highest motif proportion distributions among all models.
Another intriguing observation arises when comparing the distributions of the generated traces with those from the real data.
The result from DT-EPR displays lower entropy, signifying higher regularity in location visits, but generally has a lower motifs proportion compared to the real traces (labeled as GC in Figure~\ref{fig:demo_generation}B and C).
Next, we introduce causal interventions to the data-generation process and assess their direction and impact using mobility metrics.
Figure~\ref{fig:generation_metric} displays the distributions of mobility entropy and motifs proportion for both observational and interventional location sequences from DT-EPR. The latter are generated by implementing interventions on individuals' exploration tendencies.
Additional metrics for describing the impact of these interventions can be found in Appendix~\ref{app:metrics}.
The interventions on population-level attractiveness and individual-level preference preserve the general visitation characteristics; hence, high-level mobility metrics cannot accurately reflect their impact.

The interventions can effectively and directionally change the underlying mobility pattern, as demonstrated by the shifts in the mobility metric distributions, shown in Figure~\ref{fig:generation_metric}.
For example, with an increase in the exploration tendency $p^{new}$, individuals are encouraged to visit new locations, resulting in an increasingly higher number of transited location pairs (Appendix~\ref{app:metrics}) and leading to mobility sequences with higher entropy and lower motifs proportion (Figure~\ref{fig:generation_metric}A).
Moreover, the impact on the generated location sequences can be compared among the different interventions.
While intervening on the exploration tendency $p^{new}$ significantly alters the mobility patterns (Figure~\ref{fig:generation_metric}A), changes induced by exploration parameters $\rho$ and $\gamma$ are more nuanced and provide more fine-grained control (Figure~\ref{fig:generation_metric}B and C).
In summary, mobility metrics facilitate the description and comparison between observational and interventional location sequences, allowing us to assess the direction and strength of the interventions on mobility behavior.

\subsection{Next location prediction}\label{sec:res_prediction}

Thanks to the mobility simulation models and the causal intervention process, we can assess the performance of predictive networks when confronted with distribution shifts.
We begin by comparing the performance of LSTM and MHSA on mobility sequences synthesized by different variants of the EPR model (Table~\ref{tab:performance}), thereby uncovering the influence of generative mechanisms.
These networks are trained with location sequences from DT-EPR and evaluated on datasets obtained using the same observational parameters but different ablations of its components.
We also assess their performances on a new DT-EPR population with different initializations (e.g., for each individual a newly sampled start location, $\rho$, and $\gamma$; see \(\S\)\ref{sec:realize_model} for details of the mobility model realization).
This variant is denoted as \textit{DT-EPR re-generated} in Table~\ref{tab:performance}, with the network performance differences reflecting the effect of induced randomness in the mobility model.
When evaluated on the training population (\textit{DT-EPR trained} in Table~\ref{tab:performance}), MHSA demonstrates a stronger capability to infer the next location compared to LSTM, as indicated by the higher performance indicators Acc$@$1, F1, and MRR.
However, LSTM outperforms MHSA when confronted with all other populations, i.e., when predicting for location sequences on which the network was not trained, demonstrating superior generalization ability.
In addition, both networks exhibit the worst performance when provided with sequences obtained from the vanilla EPR model, indicating the absence of clear transition patterns in these sequences -- a property overlooked by the EPR mechanism.
As a further enhancement to the model, IPT enables the generation of location sequences with transition patterns that a predictive network can utilize to infer the next location.
Moreover, incorporating population preferences during exploration imposes essential structural patterns in location visit sequences, as evidenced by the increase in prediction performance between EPR and d-EPR, as well as IPT and DT-EPR.
These structural patterns are not apparent in the entropy and motifs proportion distributions (Figure~\ref{fig:demo_generation}), yet they contribute to increasing the accuracy of next location prediction (Table~\ref{tab:performance}).

\begin{table*}[ht!]
  \caption{Next location prediction performances for observational location sequences from EPR-like models. The mean and standard deviation across five runs with different random parameter initializations are reported.}
  \label{tab:performance}
  \centering
  \begin{tabular}{@{}ccccccc@{}}
    \toprule
    Networks              & Datasets           & Acc$@$1 (\%)          & Acc$@$5 (\%)          & Acc$@$10 (\%)          & F1 (\%)             & MRR (\%)            \\ \midrule
    \multirow{5}{*}{LSTM} & EPR                 & $1.1 \pm 0.03$  & $2.8 \pm 0.04$  & $3.9 \pm 0.02$  & $0.3 \pm 0.01$  & $2.2 \pm 0.03$  \\
                          & d-EPR               & $1.2 \pm 0.04$  & $3.2 \pm 0.1$   & $4.6 \pm 0.1$   & $0.4 \pm 0.04$  & $2.5 \pm 0.05$  \\
                          & IPT                 & $17.3 \pm 0.3$  & $25.8 \pm 0.3$  & $28.8 \pm 0.2$  & $11.2 \pm 0.1$  & $21.4 \pm 0.2$  \\
                          & DT-EPR re-generated & $25.1 \pm 0.2$  & $35.1 \pm 0.2$  & $38.5 \pm 0.2$  & $17.8 \pm 0.2$  & $30.0 \pm 0.2$  \\
                          & DT-EPR trained      & $55.0 \pm 0.1$  & $62.9 \pm 0.1$  & $64.6 \pm 0.05$ & $46.2 \pm 0.2$  & $58.6 \pm 0.1$  \\
    \midrule
    \multirow{5}{*}{MHSA} & EPR                 & $0.8 \pm 0.01$  & $2.4 \pm 0.02$  & $3.7 \pm 0.02$  & $0.3 \pm 0.01$  & $2.0 \pm 0.01$  \\
                          & d-EPR               & $0.9 \pm 0.01$  & $2.8 \pm 0.02$  & $4.2 \pm 0.04$  & $0.4 \pm 0.01$  & $2.2 \pm 0.01$  \\
                          & IPT                 & $15.0 \pm 0.2$  & $23.7 \pm 0.4$  & $27.3 \pm 0.4$  & $10.6 \pm 0.1$  & $19.3 \pm 0.2$  \\
                          & DT-EPR re-generated & $24.4 \pm 0.1$  & $33.8 \pm 0.2$  & $37.5 \pm 0.2$  & $18.7 \pm 0.04$ & $29.0 \pm 0.1$  \\
                          & DT-EPR trained      & $55.8 \pm 0.02$ & $62.7 \pm 0.04$ & $64.4 \pm 0.03$ & $47.3 \pm 0.04$ & $59.1 \pm 0.02$ \\
    \bottomrule
  \end{tabular}
\end{table*}

\subsection{Robustness of location prediction networks}

\begin{figure}[!htb]
  \centering
  \includegraphics[width=0.8\textwidth]{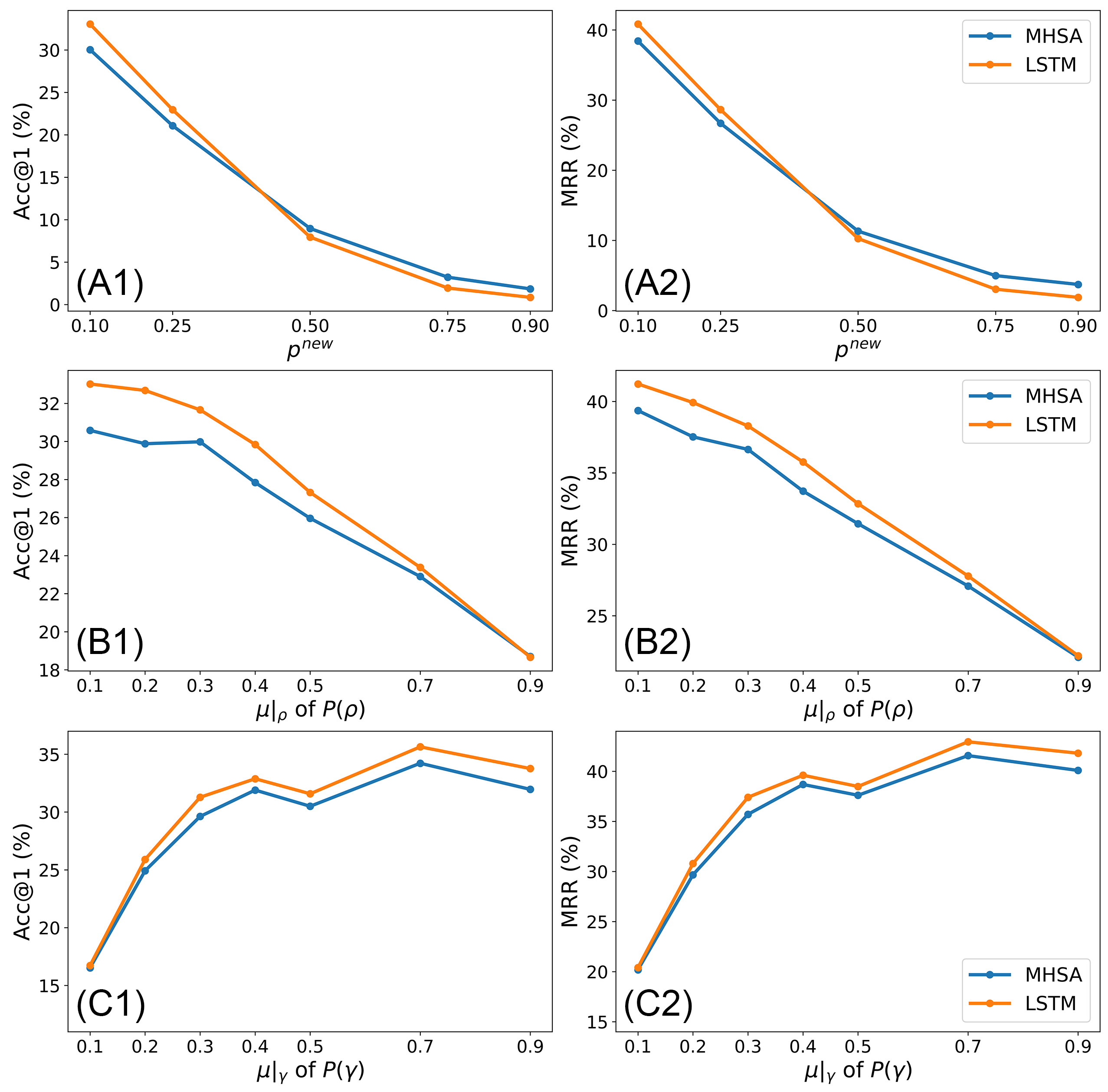}
  \caption{Next location prediction performances for interventions on individuals' exploration tendency. We show the variations in Acc$@$1 and MRR for (A) hard interventions on $p^{new}$, (B) interventions on $\rho$ by shifting $\mu\vert_{\rho}$ of $P(\rho)$, and (C) interventions on $\gamma$ by shifting $\mu\vert_{\gamma}$ of $P(\gamma)$.}
  \label{fig:interv_explore}
\end{figure}

\begin{figure}[!htbp]
  \centering
  \includegraphics[width=0.8\textwidth]{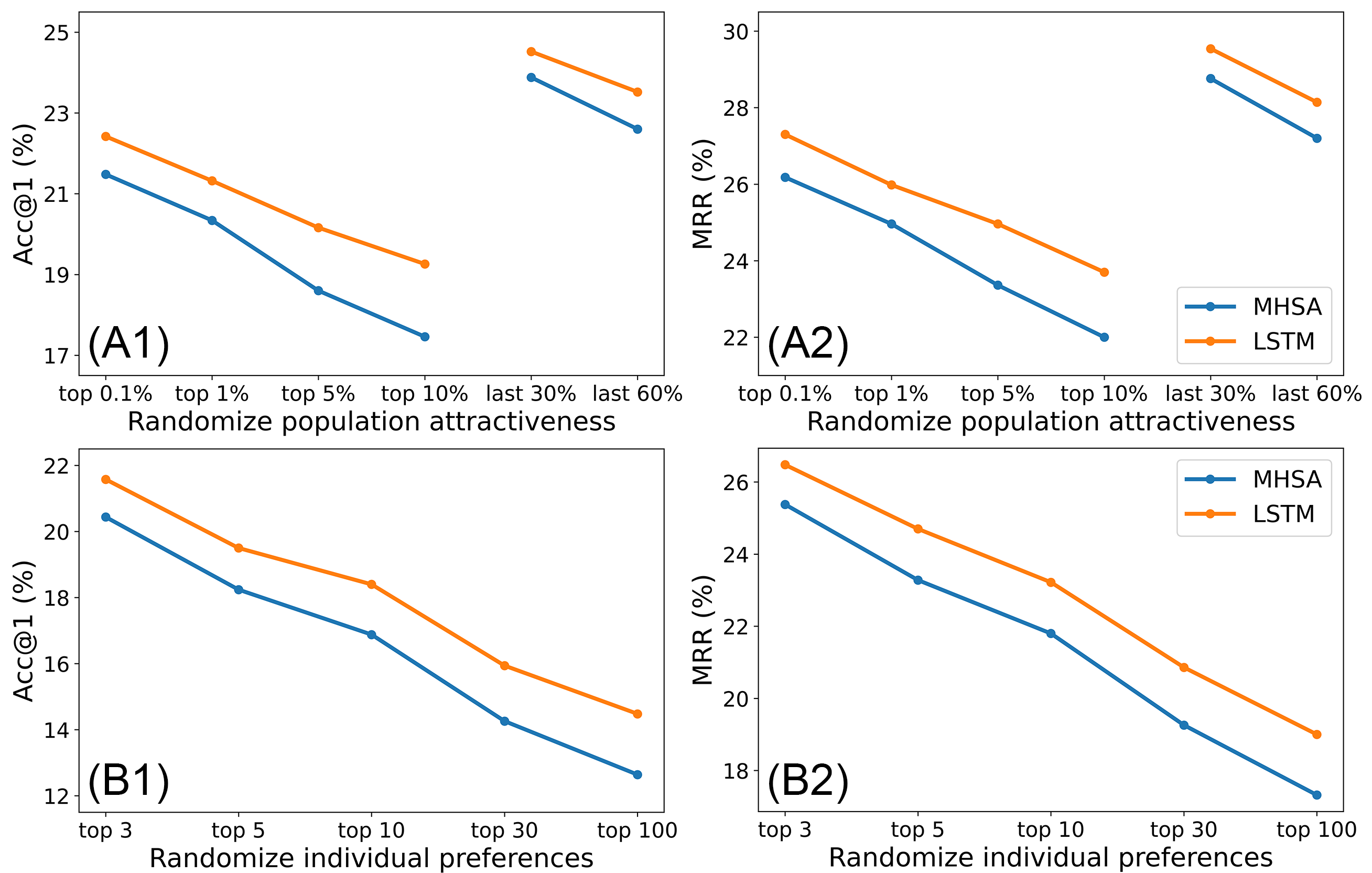}
  \caption{Next location prediction performances for interventions on population attractiveness and individual preference. We show the variations in Acc$@$1 and MRR for (A) randomizing empirical location visits of the dataset, and (B) randomizing empirical location visits for each individual.}
  \label{fig:interv_IPT_PoP}
\end{figure}

We now evaluate networks' performance using interventional mobility sequences, which reveal their robustness in OoD scenarios, i.e., when the training and testing data are not generated from the same underlying distribution.
Figure~\ref{fig:interv_explore} displays the variations in Acc$@$1 and MRR scores of the LSTM and MHSA networks for interventions on exploration tendency, and Figure~\ref{fig:interv_IPT_PoP} depicts the performance variation plot for interventions on population attractiveness and individual preference.
The complete performance results for all conducted interventions are presented in Appendix~\ref{app:performance}.
The prediction networks remain consistent with those described in the previous section, i.e., they are trained using observational location sequences from DT-EPR. Consequently, the performance metrics in OoD scenarios can be cross-compared with those presented in Table \ref{tab:performance}.
Although similar performance trends are observed for both networks, LSTM consistently outperforms MHSA in OoD settings, confirming the observation from the previous results.

The performance variations for exploration interventions (Figure~\ref{fig:interv_explore}) generally align with their strengths and directions, as measured using mobility metrics shown in Figure~\ref{fig:generation_metric}.
We also observe non-linear relations between intervention strength and prediction performance.
In particular, the hard interventions on $p^{new}$ significantly influence the prediction capability. Setting $p^{new} > 0.5$ results in performance indicators of Acc$@$1 $< 10 \%$ and MRR $< 10 \%$, suggesting that the learned location transition patterns cannot be adequately utilized.
Comparatively, interventions on $\gamma$ and $\rho$ indirectly affect $p^{new}$, which retains the diminishing exploration speed over time. As a result, influences on next location prediction are milder, e.g., the Acc$@$1 still achieves $\sim 18 \%$ with the strongest implemented interventions ($\mu\vert_{\rho} = 0.9$ and $\mu\vert_{\gamma} = 0.1$).
Moreover, we observe the prediction performances are relatively stable for $\mu\vert_{\rho} \in [0.1, 0.3]$ and $\mu\vert_{\gamma} \in [0.5, 0.9]$, even though the location sequences continue to exhibit lower mobility entropy and higher motifs proportion (Figure~\ref{fig:generation_metric}) due to decreasing exploration tendency.
This saturation suggests that even if individuals explore new locations at a lower rate, many location visit patterns are inherently stochastic and complex, making them challenging to capture by a trained network.

Interventions on population attractiveness and individual preference reveal how altering visit frequencies for specific locations affects the prediction performances, as shown in Figure~\ref{fig:interv_IPT_PoP}.
In each intervention, we consider a set of locations and randomly shuffle their empirical visitation numbers within the group; for instance, in the shuffling process, ``top 1\%'' includes the most frequently visited $1\%$ of locations across the population (Figure~\ref{fig:interv_IPT_PoP}A), and ``top 3'' considers the three most visited locations for each individual (Figure~\ref{fig:interv_IPT_PoP}B).
Both types of interventions substantially impact the ability to correctly predict the next location, with altering the number of visits separately for each individual showing a stronger influence, as evidenced by the higher drop in performance indicators.
When compared with the observational datasets (Table~\ref{tab:performance}), even changes in preference for a few most critical locations (e.g., ``top 0.1\%'' for location attractiveness or ``top 3'' for individual preferences) result in a significant prediction capability decrease.
On the contrary, intervening on a large portion of locations that are not frequently visited (i.e., ``last 30\%'' and ``last 60\%'' for location attractiveness) has minimal impact on the prediction performances.
These results emphasize the indispensable role of essential locations in shaping daily mobility and reveal their relation with the generalization ability of next location prediction networks.

\section{Discussion}\label{sec:discussion}
This study has assessed the robustness of neural networks on the next location prediction task.
We implement multiple individual mobility simulation models, on which we conduct causal interventions to synthesize mobility traces when structural changes occur.
These interventional mobility traces are then used as input for trained networks and evaluated for location prediction.
The examination of mobility simulators revealed that the vanilla EPR fails to generate sequences with realistic visitation order, while IPT tends to produce over-regular sequences by explicitly incorporating the first-order Markov dependence (Figure~\ref{fig:demo_generation}).
These findings highlight the need for further developments in individual mobility models, essential for visit order-sensitive applications, such as mobility prediction~\citep{ma_individual_2022}, autonomous vehicles scheduling~\citep{li_incorporating_2020}, and smart charging optimization~\citep{cai2022optimizing}.
Moreover, the comparison between the two sequence modeling networks, LSTM and MHSA-based, reveals their generalization characteristics. While the MHSA-based network performs better when predicting locations for the same population, LSTM demonstrates higher generalization capability when dealing with sequences sampled OoD.

Performing causal interventions on the mobility simulators enables us to control the strength and direction of mobility behavior. These interventions quantitatively reveal the behavioral impact on location predictors and have practical implications for downstream applications.
We establish a connection between exploration tendency, mobility behavior, and prediction performance using mobility metrics, providing a benchmark for comparing existing or newly developed mobility prediction networks, potentially on other datasets. This enables future studies to evaluate their population's mobility behavior and determine whether their networks perform better than those reported in this study.
Additionally, this connection aids in selecting target populations for predictive networks~\citep{solomon_analyzing_2021} and preemptively assessing the performance of demographic groups~\citep{baumann_selecting_2018}.
An improved understanding of prediction networks reveals the priorities in enhancing their behavioral robustness, which greatly benefits real-world applications such as enhanced location-based mobility services (e.g., on-demand mobility~\citep{kieu_class_2020}) and traffic management~\citep{sun_joint_2021}.

The findings from location interventions indicate that shifts in spatiotemporal preferences significantly affect prediction network performance. However, these preference shifts cannot be differentiated using high-level mobility metrics, as they alter location choices while maintaining the empirical visitation distribution. 
This observation highlights the need to develop change detection methods for identifying precise mobility behavior change points or periods~\citep{hong_clustering_2021, roy_functional_2023}.
Moreover, our results reveal that preference changes in the essential locations have a much more significant effect on the prediction network than changes in the less visited ones.
Recent studies on intra-person variability of travel behavior suggest that individuals constantly update their important locations over time~\citep{jarv_understanding_2014, alessandretti_evidence_2018}. This implies that real-world deployment of mobility prediction networks requires integrating online learning, where networks are continuously updated as new data arrives~\citep{van_lint_online_2008, jiang_deepurbanmomentum_2018}.
In the online learning framework, this study can help identify the point at which the network's performance becomes insufficient and subsequently update network parameters to adapt to evolving mobility behavior.

\section{Conclusion}\label{sec:conclusion}
Unraveling the role and impact of multifaceted mobility behavior on prediction outcomes is imperative to the real-world application of mobility prediction systems.
Here, we present a framework to examine how behavioral factors influence mobility prediction networks through causal interventions.
Using mechanistic mobility models with parameters that capture mobility behaviors, we perform causal interventions on these parameters to generate mobility traces that mirror real-world behavior variations.
Quantitative evaluation using mobility metrics demonstrates our capability to effectively and deliberately modify behaviors.
Subsequently, we evaluate these interventional traces with well-trained networks for the next location prediction task, and the resulting performance variations indicate the robustness of networks confronting domain shifts.
Our results reveal vital behavior factors affecting prediction performance, including sequential location transition patterns, the tendency to explore new locations, and location preferences at both population and individual levels.
These findings demonstrate the framework's effectiveness and pave the way for various downstream applications, including cross-comparisons among prediction networks and performance monitoring in dynamically evolving spatiotemporal scenarios.

As one of the pioneering studies to explore the causal impact of mobility behavior on individual mobility prediction, this research sheds light on several directions for future work.
The proposed framework can be applied to any mechanistic generation model abstracted as an SCM. Our interventions focus on the EPR model and its extensions, and our results are fundamentally limited by their capability to simulate activity location choices. 
Future research should investigate other mobility simulation models, such as location attractiveness~\citep{yan_universal_2017} and container~\citep{alessandretti_scales_2020}, to examine the impact of alternative behavioral mechanisms on mobility prediction.
In addition to factors that describe mobility, quantifying the causal effects of contexts (e.g., urban functions, land use, and weather situations) on mobility behavior is a promising direction. This endeavor would require a mechanistic model that explicitly formulates the relationship between contextual factors and mobility behavior, which could be inferred using causal discovery methods~\citep{runge_causal_2023}.
Last, counterfactuals offer greater causal reasoning capabilities than interventions, as outlined in the three-layer causal hierarchy of association (layer 1), intervention (layer 2), and counterfactual (layer 3) by~\cite{pearl_book_2018}. Exploring counterfactual explanations for prediction networks provides actionable suggestions that can lead to desirable alternative outcomes~\citep{xin_vision_2022}. This aspect is crucial for the real-world deployment of mobility prediction systems, even though its full potential remains to be realized.
In conclusion, we expect this study to stimulate innovative approaches that leverage causal inference to improve the interpretability and robustness of neural networks when applied to mobility analysis and prediction.


\bibliography{mybibfile}
\clearpage

\beginsupplement
\section*{Appendix}
\addcontentsline{toc}{section}{Appendix}

\subsection{Prediction network and implementation}\label{app:network}

In this section, we introduce the fundamental components of LSTM and MHSA-based networks, detailing their implementation and training processes. For specific modifications to enhance performance in the next location prediction task, we refer readers to~\citet{solomon_analyzing_2021} and~\citet{Hong_context_2023}, respectively.

\begin{figure*}[!ht]
  \centering
  \includegraphics[width=1.0\textwidth]{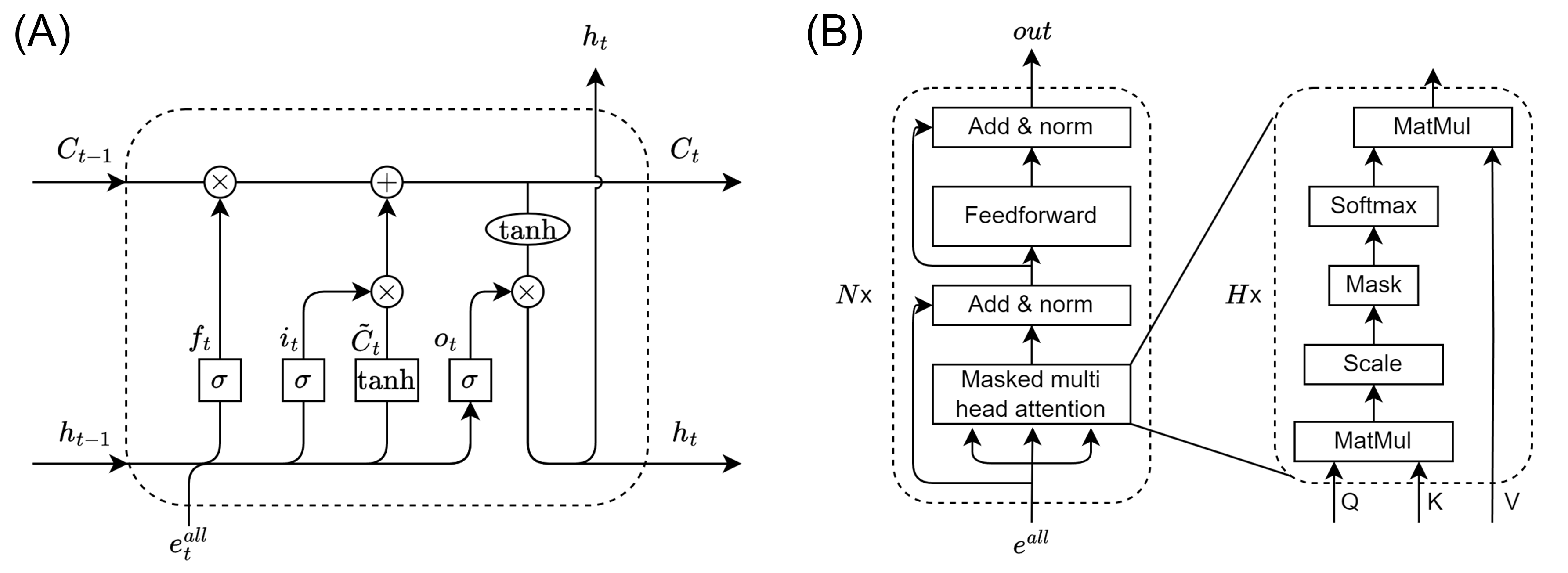}
  \caption{Network structure of (A) the LSTM network and (B) the MHSA-based network.}
  \label{fig:app_architecture}
\end{figure*}

\paragraph{\textbf{LSTM network}}

Initially proposed by \citet{hochreiter_long_1997}, LSTM improves over the vanilla recurrent neural network (RNN) for learning temporal dependencies in sequence data by addressing the vanishing gradient issue through its architecture design. 
The network processes input sequentially and, at each step $t$, utilizes three ``gates'' -- forget gate $f_{t}$, input gate $i_{t}$, and output gate $o_{t}$ -- to learn the dependencies between the current input $e^{all}_{t}$ and previous information, stored in a cell state $C_{t-1}$ and a hidden state $h_{t-1}$~(Figure~\ref{fig:app_architecture}A).
This gate-updating process can be formulated as follows:
\begin{linenomath}
  \begin{align}
    f_{t}&=\sigma( (e^{all}_{t} \oplus h_{t-1}) \cdot \boldsymbol{W}^{f} )\\
    i_{t}&=\sigma( (e^{all}_{t} \oplus h_{t-1}) \cdot \boldsymbol{W}^{i} )\\
    o_{t}&=\sigma( (e^{all}_{t} \oplus h_{t-1}) \cdot \boldsymbol{W}^{o} )
  \end{align}
\end{linenomath}
where $\oplus$ denote the concatenation operation and $\sigma$ is the sigmoid function. Then, the current cell state $C_{t}$ and hidden state $h_{t}$ can be updated as follows:
\begin{linenomath}
  \begin{align}
    {\tilde {C}}_{t}&= \tanh ( (e^{all}_{t} \oplus h_{t-1}) \cdot \boldsymbol{W}^{C} )\\
    C_{t}&=f_{t}\odot C_{t-1}+i_{t}\odot {\tilde {C}}_{t}\\
    h_{t}&=o_{t}\odot \tanh( C_{t} )
  \end{align}
\end{linenomath}
where $\odot$ denotes the Hadamard product. Finally, $C_{t}$ and $h_{t}$ are carried forward to the next step. $h_{t}$, encapsulating all information up to the current step, can also be used to infer predictions. 

\paragraph{\textbf{MHSA-based network}}

First introduced in \citet{vaswani_attention_2017}, the MHSA-based network extends the attention mechanism and has become widely adopted in sequential modeling tasks for its efficiency and ability to capture complex dependencies across the entire sequence. 
In contrast to LSTM, the MHSA-based network processes the input all at once: an embedding matrix $\boldsymbol{e}^{all}$ is constructed by stacking embedding vectors in sequence order (Figure~\ref{fig:app_architecture}B).
This input is then passed through a stack of $N$ identical layers, each containing a masked multi-head attention block and a fully connected feedforward network with two linear layers and a ReLU activation function. Residual connections~\citep{He_16} and layer normalizations~\citep{ba_layer_2016} are applied to each layer. 
The multi-head attention block uses scaled dot-product attention to compute output values based on queries and sets of key-value pairs, efficiently implemented by packing into matrices Q, K, and V~\citep{vaswani_attention_2017}: 
\begin{linenomath}
  \begin{equation}
    \text{Attention(}Q, K, V\text{)} = \text{softmax(}\frac{QK^T}{\sqrt{dim}})\cdot V
  \end{equation}
\end{linenomath}
where $dim$ is the size of the query vector. Multi-head attention is constructed by scaling the matrices Q, K, and V, and concatenating the results of $H$ attention functions:
\begin{linenomath}
  \begin{align}
    \text{MultiHead(}Q, K, V \text{)} & = (head_1 \oplus ... \oplus head_H)\cdot\boldsymbol{W}^O                                 \\
    \text{where} \quad head_i         & = \text{Attention(}Q\boldsymbol{W}^Q_i, K\boldsymbol{W}^K_i, V\boldsymbol{W}^V_i )
  \end{align}
\end{linenomath}
In each multi-head attention block, the key, value, and query matrices are identical and correspond to the output of the previous block. In addition, forward-masking operations are included to prevent attention operations from accessing information from future time steps.

\paragraph{\textbf{Implementation and training details}}
During training, we minimize Eq.~\ref{equation:loss} with the Adam optimizer \citep{Kingma_adam_2015} on batches of training data samples.
We use a batch size 256, and zeros are appended to the end of each input sequence until its length matches the longest sequence, ensuring consistent length within each batch.
The initial learning rate is set to $1e^{-3}$, and an L2 penalty of $1e^{-6}$ is applied to the network loss.
We adopt an early stopping strategy, which drops the learning rate by 0.1 if the validation loss does not decrease for 3 consecutive epochs. This early stopping process is repeated 3 times.

The tuned hyper-parameters and their search ranges are shown in Table~\ref{tab:parameter_search}. We determine the optimal set of values as the one that minimizes the network loss on the validation set of the observational traces from DT-EPR.

\begin{table}[htbp!]
  \caption{Hyper-parameter search for next location prediction networks. }
  \label{tab:parameter_search}
  \centering
  \begin{tabular}{@{}cccc@{}}
    \toprule
    Network                     & Hyper-parameter   & Search range            & Optimal value \\ \midrule
    \multirow{2}{*}{LSTM}       & Embedding dim.    & \{32, 64, 128\}         & $64$          \\
                                & Hidden dim.       & $\{64, 128, 256, 512\}$ & $128$         \\ \midrule
    \multirow{4}{*}{MHSA-based} & \#Heads $H$          & $\{2, 4, 8\}$           & $8$           \\
                                & \#Layers $N$         & $\{2, 4, 6\}$           & $4$           \\
                                & Embedding dim.    & $\{32, 64, 128\}$       & $64$          \\
                                & Feed-forward dim. & $\{64, 128, 256, 512\}$ & $256$         \\ \bottomrule
  \end{tabular}
\end{table}

\subsection{Additional metrics for describing mobility behavior}\label{app:metrics}

Apart from the mobility metrics elucidated in the main text, we introduce additional metrics to characterize mobility behavior within both observational and interventional location visitation sequences:
\begin{itemize}
  \item The radius of gyration~\citep{gonzalez_understanding_2008} measures the characteristic distance traveled by an individual and is commonly used to reflect their movement intensity in the spatial dimension. Its dataset distribution reveals the heterogeneity in user movement~\citep{xu_human_2018}.
  \item The number of transited location pairs~\citep{zhao_characteristics_2021} measures the number of links in an individual mobility graph, where locations are represented as nodes and travels between locations are represented as edges.
  \item The location visitation frequency distribution~\citep{gonzalez_understanding_2008} is known to adhere to Zipf's law, where individuals tend to spend the majority of their time at a small number of frequently visited locations. The exponent of this power law, which characterizes the shape of the distribution, reveals the frequency of important location visits and can serve as an indicator of mobility regularity.
\end{itemize}

\begin{figure*}[!ht]
  \centering
  \includegraphics[width=1.0\textwidth]{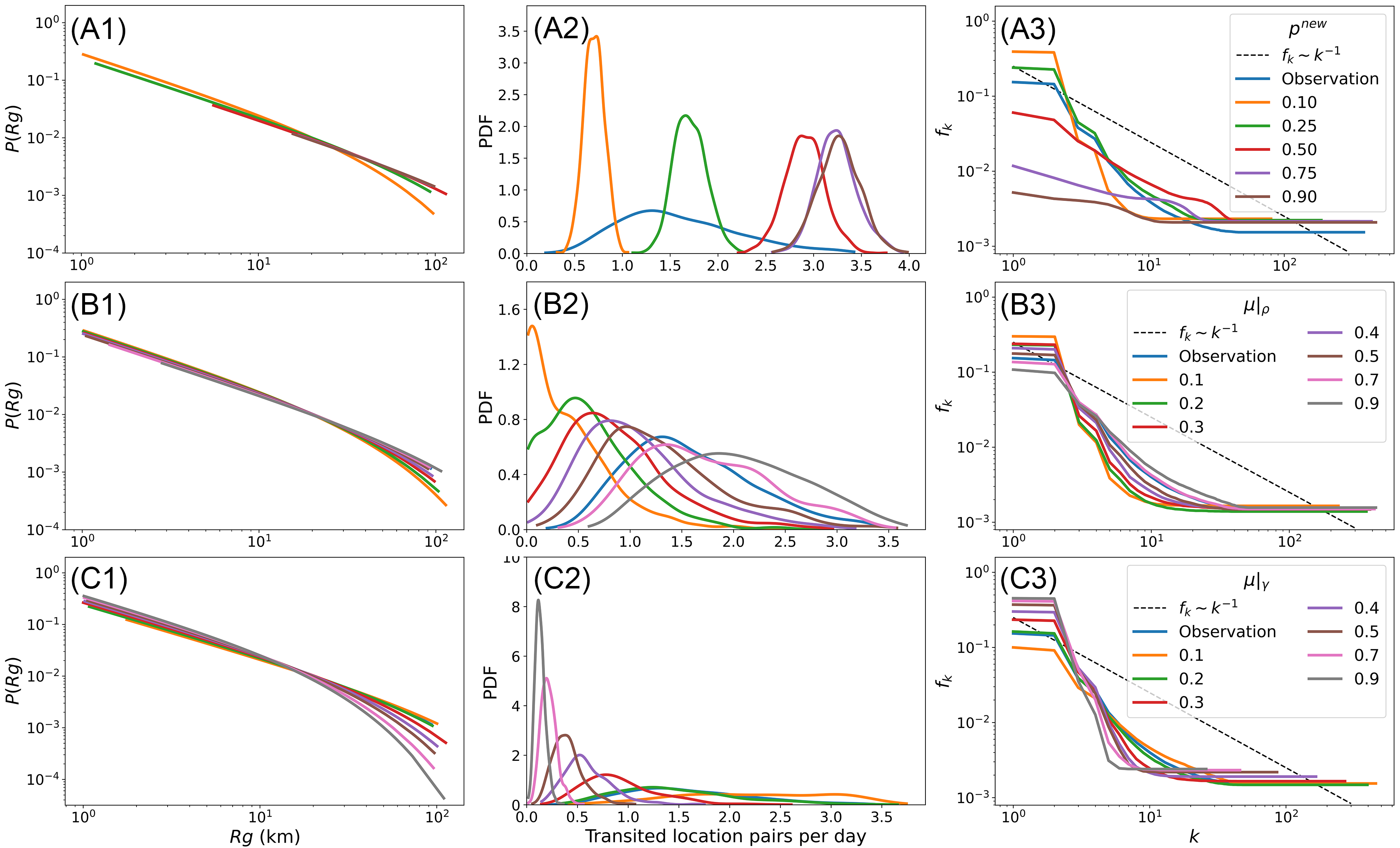}
  \caption{The radius of gyration (Rg), number of transited location pairs, and location visitation frequency distributions of observational and interventional location sequences. We show the metric distributions for (A) hard interventions on $p^{new}$, (B) interventions on $\rho$ by shifting $\mu\vert_{\rho}$ of $P(\rho)$, and (C) interventions on $\gamma$ by shifting $\mu\vert_{\gamma}$ of $P(\gamma)$.}
  \label{fig:app_generation_metrics}
\end{figure*}

Figure~\ref{fig:app_generation_metrics} illustrates alterations in mobility patterns resulting from interventions on the exploration behavior, assessed using the previously mentioned metrics.
Notably, these interventions induced changes in the population's mobility intensity, quantified by the radius of gyration. Although these interventions had a marginal impact on the prevalence of individuals traveling shorter distances, they substantially reduced the likelihood of encountering individuals with extensive travel patterns as exploration declined (lower $p^{new}$ and $\mu\vert_{\rho}$, and higher $\mu\vert_{\gamma}$).
As expected, a reduction in exploration inclination led to fewer transited location pairs per day (Figure~\ref{fig:app_generation_metrics} middle panel) and a significant decrease in the visitation frequency of the most frequently visited locations (Figure~\ref{fig:app_generation_metrics} right panel).

\subsection{Network performances on interventional location sequences}\label{app:performance}

The complete performance evaluation results for LSTM and MHSA-based networks on all simulated interventional location sequences are presented in Table~\ref{tab:lstm_performance} and Table~\ref{tab:mhsa_performance}, respectively. We report the mean and standard deviation across five network optimization runs with different random parameter initializations. 
Additionally, we note that generative mechanisms in the mobility model introduce variations in synthesized location sequences (e.g., randomness from mobility model initializations; see \(\S\)\ref{sec:res_prediction} and Table~\ref{tab:performance}), affecting mobility predictability and prediction performance. 
Therefore, we suggest focusing on the overall variation trend for interventions rather than exact performance numbers.

\begin{table*}[ht!]
  \caption{Performance of the LSTM model in predicting the next location on interventional location sequences. The mean and standard deviation across five runs with different random parameter initializations are reported.}
  \label{tab:lstm_performance}
  \centering

  \begin{tabular}{@{}ccccccc@{}}
    \toprule
    Intervention                         & Strength    & Acc$@$1 (\%)          & Acc$@$5 (\%)          & Acc$@$10 (\%)         & F1 (\%)            & MRR (\%)            \\ \midrule
    \multirow{5}{*}{$p^{new}$}           & $0.1$       & $33.0 \pm 0.6$  & $49.8 \pm 0.6$  & $55.3 \pm 0.6$  & $28.9 \pm 0.5$ & $40.8 \pm 0.6$  \\
                                         & $0.25$      & $23.0 \pm 0.3$  & $34.8 \pm 0.1$  & $38.7 \pm 0.1$  & $16.2 \pm 0.1$ & $28.6 \pm 0.3$  \\
                                         & $0.5$       & $7.9 \pm 0.1$   & $12.4 \pm 0.03$ & $14.4 \pm 0.05$ & $3.6 \pm 0.1$  & $10.3 \pm 0.04$ \\
                                         & $0.75$      & $1.9 \pm 0.1$   & $3.7 \pm 0.1$   & $4.9 \pm 0.1$   & $0.6 \pm 0.04$ & $3.0 \pm 0.1$   \\
                                         & $0.9$       & $0.9 \pm 0.04$  & $2.3 \pm 0.1$   & $3.5 \pm 0.1$   & $0.3 \pm 0.03$ & $1.9 \pm 0.1$   \\
    \midrule
    \multirow{7}{*}{$\mu\vert_{\rho}$}   & $0.1$       & $33.0 \pm 0.7$  & $50.7 \pm 0.7$  & $56.5 \pm 0.4$  & $29.2 \pm 0.6$ & $41.2 \pm 0.7$  \\
                                         & $0.2$       & $32.7 \pm 0.4$  & $48.0 \pm 0.3$  & $53.0 \pm 0.6$  & $28.3 \pm 0.3$ & $39.9 \pm 0.2$  \\
                                         & $0.3$       & $31.7 \pm 0.5$  & $45.8 \pm 0.4$  & $50.4 \pm 0.3$  & $25.9 \pm 0.4$ & $38.3 \pm 0.3$  \\
                                         & $0.4$       & $29.8 \pm 0.4$  & $42.4 \pm 0.4$  & $47.1 \pm 0.3$  & $23.5 \pm 0.4$ & $35.8 \pm 0.3$  \\
                                         & $0.5$       & $27.3 \pm 0.3$  & $38.9 \pm 0.3$  & $42.9 \pm 0.2$  & $20.5 \pm 0.3$ & $32.8 \pm 0.2$  \\
                                         & $0.7$       & $23.4 \pm 0.1$  & $32.4 \pm 0.1$  & $35.7 \pm 0.1$  & $16.2 \pm 0.1$ & $27.8 \pm 0.1$  \\
                                         & $0.9$       & $18.6 \pm 0.05$ & $25.8 \pm 0.1$  & $28.3 \pm 0.1$  & $11.5 \pm 0.2$ & $22.2 \pm 0.1$  \\
    \midrule
    \multirow{7}{*}{$\mu\vert_{\gamma}$} & $0.1$       & $16.7 \pm 0.1$  & $24.2 \pm 0.1$  & $26.9 \pm 0.1$  & $10.2 \pm 0.1$ & $20.4 \pm 0.1$  \\
                                         & $0.2$       & $25.9 \pm 0.3$  & $36.0 \pm 0.2$  & $39.4 \pm 0.3$  & $18.6 \pm 0.2$ & $30.8 \pm 0.2$  \\
                                         & $0.3$       & $31.3 \pm 0.3$  & $44.2 \pm 0.2$  & $48.2 \pm 0.6$  & $24.8 \pm 0.1$ & $37.4 \pm 0.2$  \\
                                         & $0.4$       & $32.9 \pm 0.4$  & $47.1 \pm 0.2$  & $51.7 \pm 0.2$  & $27.1 \pm 0.2$ & $39.6 \pm 0.3$  \\
                                         & $0.5$       & $31.6 \pm 0.4$  & $46.3 \pm 0.4$  & $51.2 \pm 0.3$  & $26.9 \pm 0.5$ & $38.5 \pm 0.3$  \\
                                         & $0.7$       & $35.6 \pm 0.7$  & $51.4 \pm 0.4$  & $56.7 \pm 0.7$  & $31.6 \pm 0.7$ & $42.9 \pm 0.6$  \\
                                         & $0.9$       & $33.8 \pm 0.6$  & $51.2 \pm 0.5$  & $57.0 \pm 0.9$  & $30.0 \pm 0.5$ & $41.8 \pm 0.4$  \\
    \midrule
    \multirow{6}{*}{Attractiveness}      & top $0.1\%$ & $22.4 \pm 0.2$  & $32.6 \pm 0.2$  & $35.8 \pm 0.2$  & $15.5 \pm 0.3$ & $27.3 \pm 0.1$  \\
                                         & top $1\%$   & $21.3 \pm 0.2$  & $30.9 \pm 0.1$  & $34.4 \pm 0.2$  & $14.6 \pm 0.1$ & $26.0 \pm 0.1$  \\
                                         & top $5\%$   & $20.2 \pm 0.2$  & $30.1 \pm 0.3$  & $33.4 \pm 0.1$  & $13.5 \pm 0.2$ & $25.0 \pm 0.2$  \\
                                         & top $10\%$  & $19.3 \pm 0.3$  & $28.4 \pm 0.2$  & $31.5 \pm 0.2$  & $12.6 \pm 0.2$ & $23.7 \pm 0.2$  \\
                                         & last $30\%$ & $24.5 \pm 0.3$  & $35.0 \pm 0.1$  & $38.5 \pm 0.1$  & $17.5 \pm 0.2$ & $29.5 \pm 0.2$  \\
                                         & last $60\%$ & $23.5 \pm 0.2$  & $33.1 \pm 0.3$  & $36.3 \pm 0.1$  & $16.5 \pm 0.1$ & $28.1 \pm 0.2$  \\
    \midrule
    \multirow{5}{*}{Preference}          & top $3$     & $21.6 \pm 0.2$  & $31.7 \pm 0.2$  & $35.4 \pm 0.1$  & $14.9 \pm 0.2$ & $26.5 \pm 0.2$  \\
                                         & top $5$     & $19.5 \pm 0.3$  & $30.4 \pm 0.2$  & $34.1 \pm 0.2$  & $13.5 \pm 0.2$ & $24.7 \pm 0.2$  \\
                                         & top $10$    & $18.4 \pm 0.2$  & $28.3 \pm 0.3$  & $31.8 \pm 0.4$  & $12.5 \pm 0.1$ & $23.2 \pm 0.2$  \\
                                         & top $30$    & $15.9 \pm 0.3$  & $26.2 \pm 0.2$  & $29.8 \pm 0.3$  & $10.7 \pm 0.2$ & $20.9 \pm 0.2$  \\
                                         & top $100$   & $14.5 \pm 0.5$  & $23.7 \pm 0.3$  & $27.2 \pm 0.3$  & $9.4 \pm 0.3$  & $19.0 \pm 0.3$  \\
    \bottomrule
  \end{tabular}
\end{table*}

\begin{table*}[ht!]
  \caption{Performance of the MHSA-based network in predicting the next location on interventional location sequences. The mean and standard deviation across five runs with different random parameter initializations are reported.} 
  \label{tab:mhsa_performance}
  \centering

  \begin{tabular}{@{}ccccccc@{}}
    \toprule
    Intervention                         & Strength    & Acc$@$1 (\%)          & Acc$@$5 (\%)          & Acc$@$10 (\%)         & F1 (\%)             & MRR (\%)            \\ \midrule
    \multirow{5}{*}{$p^{new}$}           & $0.1$       & $30.0 \pm 0.5$  & $48.0 \pm 0.9$  & $54.2 \pm 0.8$  & $26.6 \pm 0.3$  & $38.4 \pm 0.7$  \\
                                         & $0.25$      & $21.1 \pm 0.3$  & $32.7 \pm 0.1$  & $36.9 \pm 0.2$  & $16.1 \pm 0.1$  & $26.7 \pm 0.2$  \\
                                         & $0.5$       & $9.0 \pm 0.04$  & $13.3 \pm 0.04$ & $15.4 \pm 0.03$ & $5.3 \pm 0.04$  & $11.3 \pm 0.04$ \\
                                         & $0.75$      & $3.2 \pm 0.03$  & $6.2 \pm 0.1$   & $8.1 \pm 0.1$   & $1.5 \pm 0.04$  & $5.0 \pm 0.04$  \\
                                         & $0.9$       & $1.8 \pm 0.04$  & $4.9 \pm 0.1$   & $7.0 \pm 0.1$   & $0.9 \pm 0.04$  & $3.7 \pm 0.1$   \\
    \midrule
    \multirow{7}{*}{$\mu\vert_{\rho}$}   & $0.1$       & $30.6 \pm 0.4$  & $49.1 \pm 0.6$  & $55.8 \pm 0.7$  & $27.8 \pm 0.4$  & $39.4 \pm 0.5$  \\
                                         & $0.2$       & $29.9 \pm 0.2$  & $45.9 \pm 0.3$  & $51.9 \pm 0.6$  & $26.4 \pm 0.02$ & $37.5 \pm 0.2$  \\
                                         & $0.3$       & $30.0 \pm 0.7$  & $44.0 \pm 0.5$  & $48.9 \pm 0.6$  & $25.4 \pm 0.6$  & $36.7 \pm 0.7$  \\
                                         & $0.4$       & $27.8 \pm 0.2$  & $40.0 \pm 0.3$  & $45.2 \pm 0.4$  & $22.8 \pm 0.2$  & $33.7 \pm 0.2$  \\
                                         & $0.5$       & $25.9 \pm 0.3$  & $37.3 \pm 0.5$  & $41.6 \pm 0.4$  & $20.5 \pm 0.2$  & $31.5 \pm 0.3$  \\
                                         & $0.7$       & $22.9 \pm 0.2$  & $31.3 \pm 0.3$  & $34.9 \pm 0.2$  & $17.2 \pm 0.2$  & $27.1 \pm 0.2$  \\
                                         & $0.9$       & $18.7 \pm 0.05$ & $25.4 \pm 0.2$  & $28.1 \pm 0.1$  & $13.0 \pm 0.04$ & $22.1 \pm 0.1$  \\
    \midrule
    \multirow{7}{*}{$\mu\vert_{\gamma}$} & $0.1$       & $16.5 \pm 0.2$  & $23.8 \pm 0.1$  & $26.8 \pm 0.1$  & $11.4 \pm 0.2$  & $20.2 \pm 0.2$  \\
                                         & $0.2$       & $24.9 \pm 0.3$  & $34.6 \pm 0.3$  & $38.3 \pm 0.3$  & $19.2 \pm 0.2$  & $29.6 \pm 0.3$  \\
                                         & $0.3$       & $29.6 \pm 0.5$  & $42.4 \pm 0.1$  & $46.9 \pm 0.1$  & $24.4 \pm 0.5$  & $35.7 \pm 0.3$  \\
                                         & $0.4$       & $31.9 \pm 0.3$  & $46.0 \pm 0.4$  & $51.1 \pm 0.3$  & $27.3 \pm 0.2$  & $38.7 \pm 0.3$  \\
                                         & $0.5$       & $30.5 \pm 0.3$  & $45.3 \pm 0.4$  & $50.7 \pm 0.4$  & $26.8 \pm 0.2$  & $37.6 \pm 0.4$  \\
                                         & $0.7$       & $34.2 \pm 0.8$  & $49.4 \pm 0.7$  & $55.3 \pm 1.2$  & $31.0 \pm 0.7$  & $41.6 \pm 0.8$  \\
                                         & $0.9$       & $32.0 \pm 0.4$  & $49.0 \pm 0.6$  & $55.5 \pm 0.3$  & $29.0 \pm 0.4$  & $40.1 \pm 0.3$  \\
    \midrule
    \multirow{6}{*}{Attractiveness}      & top $0.1\%$ & $21.5 \pm 0.2$  & $31.2 \pm 0.4$  & $34.8 \pm 0.3$  & $16.0 \pm 0.1$  & $26.2 \pm 0.2$  \\
                                         & top $1\%$   & $20.3 \pm 0.2$  & $29.6 \pm 0.3$  & $33.2 \pm 0.4$  & $15.1 \pm 0.2$  & $24.9 \pm 0.2$  \\
                                         & top $5\%$   & $18.6 \pm 0.1$  & $28.3 \pm 0.3$  & $32.0 \pm 0.3$  & $13.5 \pm 0.1$  & $23.4 \pm 0.1$  \\
                                         & top $10\%$  & $17.4 \pm 0.2$  & $26.6 \pm 0.1$  & $30.4 \pm 0.1$  & $12.5 \pm 0.1$  & $22.0 \pm 0.1$  \\
                                         & last $30\%$ & $23.9 \pm 0.2$  & $34.1 \pm 0.2$  & $37.7 \pm 0.2$  & $18.3 \pm 0.1$  & $28.8 \pm 0.2$  \\
                                         & last $60\%$ & $22.6 \pm 0.3$  & $32.0 \pm 0.3$  & $35.6 \pm 0.3$  & $17.1 \pm 0.2$  & $27.2 \pm 0.3$  \\
    \midrule
    \multirow{5}{*}{Preference}          & top $3$     & $20.4 \pm 0.1$  & $30.4 \pm 0.3$  & $34.3 \pm 0.3$  & $15.4 \pm 0.1$  & $25.4 \pm 0.2$  \\
                                         & top $5$     & $18.3 \pm 0.3$  & $28.5 \pm 0.3$  & $32.4 \pm 0.2$  & $13.7 \pm 0.2$  & $23.3 \pm 0.3$  \\
                                         & top $10$    & $16.9 \pm 0.1$  & $26.7 \pm 0.2$  & $30.6 \pm 0.2$  & $12.3 \pm 0.2$  & $21.8 \pm 0.1$  \\
                                         & top $30$    & $14.3 \pm 0.2$  & $24.5 \pm 0.3$  & $28.5 \pm 0.1$  & $10.5 \pm 0.2$  & $19.3 \pm 0.2$  \\
                                         & top $100$   & $12.6 \pm 0.3$  & $22.1 \pm 0.3$  & $25.9 \pm 0.3$  & $9.2 \pm 0.1$   & $17.3 \pm 0.3$  \\
    \bottomrule
  \end{tabular}
\end{table*}

\end{document}